\documentstyle[twocolumn,seceq]{jpsj}
\def\tsum{\mathop{{\rm T}_{\Sigma}}}
\newcommand{\myref}[1]{(\ref{#1})}
\newcommand{\vecvar}[1]{\mbox{\boldmath$#1$}}

\newcommand{\nknkd}[4]{(\dalpha_{\sigma(#1)})^{\lambda_{#1}}
\cdots^{\stackrel{\scriptstyle #2}{\vee}}\!
\cdots^{\stackrel{\scriptstyle #3}{\vee}}\!\cdots
(\dalpha_{\sigma(#4)})^{\lambda_{#4}}}
\newcommand{\nknkdp}[4]{(\dalpha_{\sigma^{\prime}(#1)})^{\lambda_{#1}}
\cdots^{\stackrel{\scriptstyle #2}{\vee}}\!
\cdots^{\stackrel{\scriptstyle #3}{\vee}}\!\cdots
(\dalpha_{\sigma^{\prime}(#4)})^{\lambda_{#4}}}
\newcommand{\nksp}[3]{(\dalpha_{\sigma^{\prime}(#1)})^{\lambda_{#1}}
\cdots^{\stackrel{\scriptstyle #2}{\vee}}\!
\cdots(\dalpha_{\sigma^{\prime}(#3)})^{\lambda_{#3}}}
\newcommand{\nksd}[3]{(\dalpha_{\sigma(#1)})^{\lambda_{#1}}
\cdots^{\stackrel{\scriptstyle #2}{\vee}}\!
\!\cdots(\dalpha_{\sigma(#3)})^{\lambda_{#3}}}
\newcommand{\nkdr}[4]{\dalphar{#4}_{#1}
\cdots^{\stackrel{\scriptstyle #2}{\vee}}\!\cdots\dalphar{#4}_{#3}}
\newcommand{\dalzero}[1]{\Bigr|_{\dalpha_{#1}\sim 0}}
\newcommand{\bpr}[1]{b^{+[#1]}}
\newcommand{\dalphar}[1]{\alpha^{\dagger[#1]}}

\def\runtitle{Hi-Jack Polynomials and Quantum Calogero Model}
\def\runauthor{Hideaki {\sc Ujino} and Miki {\sc Wadati}}
\def\Define{\mathop{\stackrel{\rm def}{=}}}
\def\Hchat{{{\hat H}_{\rm C}}}
\def\Hstilde{{{\tilde H}_{\rm S}}}
\def\Hc{{H_{\rm C}}}
\def\Hs{{H_{\rm S}}}
\def\Lcheck{{\check L}}
\def\Qcheck{{\check Q}}
\def\Lm{{L^{-}}}
\def\Lp{{L^{+}}}
\def\Lmhat{{{\hat L}^{-}}}
\def\Lphat{{{\hat L}^{+}}}
\def\phighat{{{\hat \phi}_{\rm g}}}

\def\psigtilde{{{\tilde \psi}_{\rm g}}}
\def\dalpha{\alpha^{\dagger}}

\def\Lcal{{\tilde {\cal L}}}
\def\ledo{{\stackrel{\rm D}{\leq}}}
\def\Sym{\Bigr|_{\rm Sym}}
\def\bp{b^{+}}
\def\Mae{\mbox{\hspace*{-20pt}}}
\def\Cmae{\mbox{\hspace*{-10pt}}}
\def\Usr{\mbox{\hspace*{20pt}}}
\def\Cusr{\mbox{\hspace*{10pt}}}
\def\Rest{\Bigr|_{\dalphar{N+1}_{1,k+2,\cdots,N}\sim 0}}
\def\Listone{Polychronakos_1,Ujino_1,Ujino_2,Ujino_3,Ujino_4}
\title{Rodrigues Formula for Hi-Jack Symmetric Polynomials Associated
with the Quantum Calogero Model}
\author{Hideaki
{\sc Ujino}\footnote{E-mail: ujino@monet.phys.s.u-tokyo.ac.jp}
and Miki {\sc Wadati}}
\inst{Department of Physics, Graduate School of Science,
University of Tokyo,\\
Hongo 7--3--1, Bunkyo-ku, Tokyo 113}
\recdate{April 12, 1996}
\abst{The Hi-Jack symmetric polynomials, which are associated with the
simultaneous eigenstates for the first and second conserved operators of
the quantum Calogero model, are studied. Using the algebraic properties of
the Dunkl operators for the model, we derive the Rodrigues formula for
the Hi-Jack symmetric polynomials. Some properties of the Hi-Jack
polynomials and  the relationships with the Jack symmetric polynomials
and with the basis given by the QISM approach are presented.
The Hi-Jack symmetric polynomials are strong candidates
for the orthogonal basis of the quantum Calogero model.}
\kword{quantum Calogero model, Hi-Jack symmetric polynomials,
Rodrigues formula,
quantum inverse scattering method (QISM), Dunkl operator}
\begin{document}
\newtheorem{definition}{Definition}[section]
\newtheorem{proposition}{Proposition}[section]
\newtheorem{corollary}{Corollary}[section]
\newtheorem{lemma}{Lemma}[section]
\sloppy
\maketitle
\footnotetext{cond-mat/9609041, UT/Wu012, April, 1996.
Published in J. Phys. Soc. Jpn. {\bf 65} (1996) 2423--2439}
\footnotetext{JPSJ online service:
{\tt http://wwwsoc.nacsis.ac.jp/jps/jpsj/}}

\section{Introduction}\label{sec:introduction}
Exact solutions for the Schr\"odinger equations have provided important
significance in physics and mathematical physics.
Most of us have studied the Laguerre polynomials and the
spherical harmonics in the theory of the hydrogen atom,
and the Hermite polynomials and their Rodrigues formula in the theory of
the quantum harmonic oscillator. The former is also a good example that
shows the role of conserved operators in quantum mechanics. The hydrogen
atom has three, independent and mutually commuting conserved operators,
namely, the Hamiltonian, the total angular momentum and its $z$-axis
component. The simultaneous eigenstates for the three conserved operators
give the orthogonal basis of the hydrogen atom.
A classical system with a set of independent and mutually Poisson commuting
(involutive) conserved quantities whose number of elements is the same as
the degrees of freedom of the system can be integrated by quadrature.
This is guaranteed by the Liouville theorem.
Such a system is called the completely integrable system.
Quantum systems with enough number of such conserved operators are
analogously called quantum integrable systems.
The hydrogen atom is a simple example of the quantum integrable system.

Among the various quantum integrable systems, one-dimensional quantum
many-body systems with inverse-square long-range interactions are now
attracting much interests of theoretical physicists.
Of the various integrable inverse-square-interaction models, the quantum
Calogero model~\cite{Calogero_1} has the longest history. Its Hamiltonian
is expressed as
\begin{equation}
  \Mae \Hchat = 
  \frac{1}{2}\sum_{j=1}^{N}\bigl(p_{j}^{2}+\omega^{2}x_{j}^{2}\bigr)
  +\frac{1}{2}\sum_{\stackrel{\scriptstyle j,k=1}{j\neq k}}^{N}
  \frac{a^{2}-a}{(x_{j}-x_{k})^{2}},
  \label{eqn:Calogero_model}
\end{equation}
where the constants $N$, $a$ and $\omega$ are the particle number,
the coupling parameter and the strength of the external harmonic well
respectively.
The momentum operator $p_{j}$ is given by a differential operator,
$p_{j} = -{\rm i}\frac{\partial}{\partial x_{j}}$.
This model is known to be a quantum integrable system in the sense that
it has sufficient number of independent and mutually commuting conserved
operators.~\cite{Polychronakos_1,Ujino_1,Ujino_2,Ujino_3,Ujino_4}
On the other hand, the Sutherland model~\cite{Sutherland_1},
which is a one-dimensional quantum integrable system with
inverse-sine-square interactions,
\begin{equation}
  \Hstilde=\frac{1}{2}\sum_{j=1}^{N}p_{j}^{2}
  +\frac{1}{2}\sum_{\stackrel{\scriptstyle j,k=1}{j\neq k}}^{N}
  \frac{a^{2}-a}{\sin^{2}(x_{j}-x_{k})},
  \label{eqn:Sutherland_model}
\end{equation}
has been thoroughly investigated and its orthogonal basis is known to
consist of the Jack symmetric polynomials.~\cite{Stanley_1,Macdonald_1}
The quantum inverse scattering method and the Dunkl operator
(exchange operator) formalism showed that these two models 
share the same algebraic 
structure.~\cite{\Listone,Bernard_1,Bernard_2,Hikami_1,Hikami_2}
The fact strongly suggests some similarities in the structures of their
Hilbert spaces.
In order to clarify this problem,
we shall apply a naive approach that we use in the study of
the hydrogen atom to the quantum Calogero model and study the deformed
multivariable extension of the Hermite polynomials, namely, the Hi-Jack
(hidden-Jack) symmetric polynomials.~\cite{Ujino_5,Ujino_6}

The Jack symmetric polynomials are uniquely determined
by three properties.
First, they are the eigenfunctions of the differential operator
that is derived from the Hamiltonian of the Sutherland model.
Second, they possess triangular expansions in monomial symmetric
polynomials with respect to the dominance ordering.
And last, they are properly normalized.~\cite{Stanley_1,Macdonald_1}
For detail, see \myref{eqn:Jack_definition}.
Quite recently, Lapointe and Vinet discovered the Rodrigues formula for
the Jack symmetric polynomials using the Dunkl operator for the Sutherland
model.~\cite{Lapointe_1} 
In our previous letter,~\cite{Ujino_6} we extended
their results to the quantum Calogero model and gave the Rodrigues formula
for the Hi-Jack symmetric polynomials.
Since the Dunkl operators for
the Sutherland model and the Calogero model share the same algebraic
relations, all the results in our letter was obtained by translating the
corresponding results in ref. \citen{Lapointe_1}. However, some parts of
their proofs rely on the explicit expressions of the Dunkl operators
for the Sutherland model and cannot be translated into the Dunkl
operators  for the Calogero model. We noticed that the proof of the
Rodrigues formula only needs the commutator algebra among the Dunkl
operators. One of the aims of this paper is to present a proof of
the Rodrigues formula for
the Hi-Jack symmetric polynomials.
Another aim is to investigate the Hi-Jack symmetric polynomials
and to compare them with the basis of the model that was given by
QISM~\cite{Ujino_1,Ujino_2} and the Dunkl
operators.~\cite{Brink_1,Brink_2}
The algebraic construction of the eigenstates for
the Hamiltonian (the first conserved operator) of the quantum
Calogero model
has already been given. Thus the Hi-Jack symmetric polynomials
must be
linear combinations of them. We shall specify the linear
combinations that
relate the Hi-Jack polynomials and the eigenstates
of the Hamiltonian
given before.~\cite{Ujino_1,Ujino_2,Brink_1,Brink_2}
We also want to see the relationships and similarities 
between the Jack polynomials and the Hi-Jack polynomials.
Though we have
presented the Rodrigues formula for the Hi-Jack ``polynomials,''
but we  have not explicitly shown that they are really polynomials.
We shall present a clear answer to these questions.

The outline is as follows. In \S\ref{sec:preparations},
we summarize and
reformulate the results of QISM and Dunkl operators' approach to
the Calogero and Sutherland models.
The Jack symmetric polynomials are also introduced.
In \S\ref{sec:Hi-Jack}, we present the Rodrigues formula
for the Hi-Jack
symmetric polynomials and introduce some propositions
that guarantee the
results. Some properties of the Hi-Jack polynomials are
also presented.
In \S\ref{sec:proofs}, we prove the propositions.
And in the final
section, we give a brief summary and discuss future problems.

\section{Models and Formulations}\label{sec:preparations}
In our derivation of the Rodrigues formula for the Hi-Jack
symmetric polynomials, we do many computations involving the Dunkl
operators for the Calogero model. We also compare the Hi-Jack
polynomials
with the Jack polynomials and with the basis of the Calogero model
given by the QISM approach.
Thus we need a summary of the Calogero model, the Sutherland model,
the QISM approach and the Dunkl operator formalism.

First, we reformulate the QISM and the Dunkl operators for the
Calogero model.
The Lax matrices for the $N$-body system are given by $N\times N$
operator-valued matrices. 
To express them, we have to introduce two operator-valued matrices:
\begin{subequations}
  \begin{eqnarray}
    \Lcheck_{jk}
    & = & {\rm i}p_{j}\delta_{jk}-a(1-\delta_{jk})\frac{1}{x_{j}-x_{k}},
    \label{eqn:L-check_matrix}\\
    \Qcheck_{jk} & = & x_{j}\delta_{jk},
    \label{eqn:Q-check_matrix}
  \end{eqnarray}
  \label{eqn:LQ-check_pair}
\end{subequations}
where $j,k=1,2,\cdots,N$.
The above $\Lcheck$-matrix is for the (rational) Calogero-Moser model
whose
Hamiltonian is obtained by taking $\omega=0$ of the Calogero model
or the
rational limit of the Sutherland model.
The Lax matrices for the Calogero model are
\begin{subequations}
  \begin{eqnarray}
    \Lmhat & = & \Lcheck + \omega \Qcheck,
    \label{eqn:L_minus_hat_matrix}\\
    \Lphat & = & -\frac{1}{2\omega}(\Lcheck - \omega \Qcheck).
    \label{eqn:L_plus_hat_matrix}
  \end{eqnarray}
  \label{eqn:L_plus_minus_hat_pair}
\end{subequations}
In eqs. \myref{eqn:LQ-check_pair} --
\myref{eqn:L_plus_minus_hat_pair} above, we introduced unusual
normalizations and accent marks for convenience of later
discussions.
Then the Hamiltonian \myref{eqn:Calogero_model} is expressed
by the above
Lax matrices as
\begin{eqnarray}
  \Hchat & = & \omega\tsum\Lphat\Lmhat + \frac{1}{2}
  N\omega\bigl(Na + (1-a)\bigr)
  \nonumber\\
  & = &  \omega\tsum\Lphat\Lmhat + E_{\rm g},
  \label{eqn:Calogero_in_L_plus_minus}
\end{eqnarray}
where $\tsum$ denotes a sum over all the matrix elements,
$\tsum A=\sum_{i,j}A_{ij}$, and $E_{\rm g}$ is the ground
state energy.
Note that the first term of the r.h.s of eq.
\myref{eqn:Calogero_in_L_plus_minus} is a nonnegative
hermitian operator.
Thus the ground state is the solution of the following equations,
\begin{eqnarray}
  & & \sum_{k=1}^{N}\Lmhat_{jk}\phighat = 0,\mbox{ for }j=1,2,\cdots,N
  \nonumber\\
  & & \Rightarrow \Hchat\phighat = E_{\rm g}\phighat.
  \label{eqn:Calogero_ground_state_equation}
\end{eqnarray}
The ground state wave function is the real Laughlin wave function:
\begin{equation}
  \phighat = \prod_{1\leq j<k\leq N}|x_{j}-x_{k}|^{a}
  \exp\bigl(-\frac{1}{2}\omega\sum_{j=1}^{N}x_{j}^{2}\bigr).
  \label{eqn:Real_Laughlin}
\end{equation}
A short note might be in order.
The phase of the difference product of the above
real Laughlin wave function, which determines the statistics of
the particles,
or in other words, the symmetry of all the eigenstates, can
be arbitrary.
We can assign any phase factor to all the exchanges of particles.
However,
we must introduce a phase factor to the definition of
the Dunkl
operators.~\cite{Ujino_4}
To avoid unnecessary complexity, we fix the phase unity.

The eigenstate of the Calogero model is fatorized into an
inhomogeneous symmetric polynomial and the ground state wave
function.
For convenience of investigations on the inhomogeneous
symmetric polynomials,
we redefine the Lax matrices \myref{eqn:L_plus_minus_hat_pair}
by the following
similarity transformation:
\begin{subequations}
  \begin{eqnarray}
    \Lm & = & \phighat^{-1}\Lmhat\phighat,
    \label{eqn:L_minus}\\
    \Lp & = & \phighat^{-1}\Lphat\phighat.
    \label{eqn:L_plus}
  \end{eqnarray}
  \label{eqn:L_plus_minus_pair}
\end{subequations}
Any operator with a hat, ${\hat {\cal O}}$, is related to
an operator ${\cal O}$ by the similarity transformation
using the ground
state wave function $\phighat$,
\begin{subequations}
  \begin{eqnarray}
    {\cal O} & = & \phighat^{-1}{\hat {\cal O}}\phighat,
    \label{eqn:Harmonic_similar_transformation}\\
    {\hat {\cal O}} & = & \phighat{\cal O}\phighat^{-1}.
    \label{eqn:Inverse_harmonic_similar_transformation}
  \end{eqnarray}
  \label{eqn:Harmonic_similar}
\end{subequations}
A set of mutually commuting conserved operators 
of the Calogero model $\{I_{n}|n=1,2,\cdots,N\}$ is given by
\begin{equation}
  I_{n} = \tsum (\Lp\Lm)^{n}.
  \label{eqn:Calogero_conserved_operators}
\end{equation}
The Hamiltonian $\Hc$ is equal to $\omega I_{1}+E_{\rm g}$. 
We regard the first conserved operator $I_{1}$ as
the Hamiltonian of the
Calogero model.
The Heisenberg equations for the $\Lm$ and $\Lp$
matrices are expressed
in the forms of the Lax equation.~\cite{Ujino_1} Moreover,
we have more general relations for a class of operators,
\begin{equation}
  V_{p}^{m} = \tsum[(\Lm)^{m}(\Lp)^{p}]_{\rm W},
  \label{eqn:Calogero_Weyl_ordered_operators}
\end{equation}
where the subscript W means the Weyl ordered product.
The class of operators naturally includes the Hamiltonian,
$\Hc=\omega V_{1}^{1}$.
The generalized Lax equations are
\begin{subequations}
  \begin{eqnarray}
    \mbox{\hspace*{-20pt}}
    \bigl[V_{p}^{m},\Lm\bigr] & = & \bigl[\Lm,Z_{p}^{m}\bigr]
    -p[(\Lm)^{m}(\Lp)^{p-1}]_{\rm W},
    \label{eqn:Calogero_generalized_Lax_minus}\\
    \mbox{\hspace*{-20pt}}
    \bigl[V_{p}^{m},\Lp\bigr] & = & \bigl[\Lp,Z_{p}^{m}\bigr]
    +m[(\Lm)^{m-1}(\Lp)^{p}]_{\rm W},
    \label{eqn:Calogero_generalized_Lax_plus}
  \end{eqnarray}
  \label{eqn:Calogero_generalized_Lax}
\end{subequations}
where the symbol $Z_{p}^{m}$ is an $N\times N$ operator-valued
matrix that satisfies the sum-to-zero condition:
\begin{equation}
  \sum_{j=1}^{N}(Z_{p}^{m})_{jk} = 
  \sum_{j=1}^{N}(Z_{p}^{m})_{kj} = 0,\;\;\;
  \mbox{for }k=1,2,\cdots,N.
  \label{eqn:Calogero_sum-to-zero}
\end{equation}
The generalized Lax equations exhibit that the operators
\myref{eqn:Calogero_Weyl_ordered_operators} satisfy
the commutation relations
of $W$-algebra.~\cite{Ujino_2,Ujino_3}

The operators with $m=0$ are important
in the construction of the eigenstates of the Hamiltonian,
because they satisfy
\begin{equation}
  \bigl[I_{1},V_{p}^{0}\bigr]=pV_{p}^{0},\;\;\;\mbox{for }
  p=1,2,\cdots.
  \label{eqn:Calogero_creation_operators}
\end{equation}
Thus the operators $V_{p}^{0}$ play the same role as the creation
operator
in the theory of the quantum harmonic oscillator. We call these
mutually
commuting operators $V_{p}^{0}$ power sum creation operators,
whose meaning will be clear later.

Successive operations of the power sum creation operators generate
all the eigenstates of the Hamiltonian, which are
labeled by the Young tableaux. The Young tableau $\lambda$ is
a non-increasing
sequence of $N$ nonnegative integers:
\begin{equation}
  \lambda = \{\lambda_{1}\geq\lambda_{2}\geq\cdots\geq\lambda_{N}\geq 0\}.
  \label{eqn:Young_tableau}
\end{equation}
Then the polynomial part of the excited state $\phi_{\lambda}$
is given by~\cite{Ujino_1,Ujino_2}
\begin{eqnarray}
  \phi_{\lambda} & = & (V_{N}^{0})^{\lambda_{N}}
  (V_{N-1}^{0})^{\lambda_{N-1}-\lambda_{N}}\cdots
  (V_{1}^{0})^{\lambda_{1}-\lambda_{2}}\cdot 1
  \nonumber\\
  & = & \prod_{p=1}^{N}(V_{p}^{0})^{\lambda_{k}-\lambda_{k+1}}
  \cdot 1,
  \label{eqn:Eigenstates_QISM}
\end{eqnarray}
where $\lambda_{N+1} = 0$
and the eigenvalue for the first conserved operator is
\begin{eqnarray}
  I_{1}\phi_{\lambda} & = &
  \sum_{k=1}^{N}\lambda_{k}\phi_{\lambda}\nonumber\\
  & = & E_{1}(\lambda)\phi_{\lambda}.
  \label{eqn:Eigenvalue_1st}
\end{eqnarray}
It is not trivial that $\phi_{\lambda}$ is indeed 
an inhomogeneous symmetric polynomial.
We shall prove it in \S\ref{sec:proofs}.
Note that the eigenstate of the original Hamiltonian $\Hchat$
\myref{eqn:Calogero_model} and its eigenvalue are
${\hat \phi}_{\lambda}=\phighat\phi_{\lambda}$ and
$\omega E_{1}(\lambda)+E_{\rm g}$.
These eigenstates give the complete set of the eigenstates. However,
they are not orthogonal because of the remaining large degeneracy.

Using the Dunkl operator formalism, we can do an analogous
investigation on the
Calogero model. The Dunkl operators for the model are
\begin{subequations}
  \begin{eqnarray}
    \alpha_{l} & = & {\rm i}\bigl(p_{l}
    +{\rm i}a\sum_{\stackrel{\scriptstyle k=1}{k\neq l}}^{N}
    \frac{1}{x_{l}-x_{k}}(K_{lk}-1)\bigr),
    \label{eqn:Annihilation}\\
    \dalpha_{l} & = & -\frac{\rm i}{2\omega}\bigl(p_{l}
    +{\rm i}a\sum_{\stackrel{\scriptstyle k=1}{k\neq l}}^{N}
    \frac{1}{x_{l}-x_{k}}(K_{lk}-1) +2{\rm i}\omega x_{l}\bigr),
    \nonumber\\
    \label{eqn:Creation}\\
    d_{l} & = & \dalpha_{l}\alpha_{l},
    \label{eqn:d_operator}
  \end{eqnarray}
  \label{eqn:Dunkl}
\end{subequations}
where $K_{lk}$ is the coordinate exchange operator.
The operator $K_{lk}$ has the properties
\begin{eqnarray}
  & & K_{lk} = K_{kl}, \;\; (K_{lk})^{2} = 1, \;\;
  K_{lk}^{\dagger}=K_{lk},\;\;
  K_{lk}\cdot 1 = 1,\nonumber\\
  & & K_{lk}A_{l}=A_{k}K_{lk}, \;\; K_{lk}A_{j} = A_{j}K_{lk},\;\;
  \mbox{for }j\neq l,k,
  \label{eqn:Exchange_operator}
\end{eqnarray}
where $A_{j}$ is either a partial differential operator 
$\frac{\partial}{\partial x_{j}}$ (or equivalently,
a momentum operator $p_{j}$), a particle coordinate
$x_{j}$ or coordinate exchange operators $K_{jk}$, $k=1,2,\cdots,N$, 
$k\neq j$.
Note that the action on the ground state of the above Dunkl
operators has
already been
removed by the similarity transformation \myref{eqn:Harmonic_similar}.
The Dunkl operators satisfy the relations,
\begin{subequations}
  \begin{eqnarray}
    & & [\alpha_{l},\alpha_{m}] = 0,\;\;\;
    [\dalpha_{l},\dalpha_{m}] = 0,
    \label{eqn:Commutator_harmonic_Dunkl_0}\\
    & & [\alpha_{l},\dalpha_{m}]
    \nonumber\\
    & & \;\;= \delta_{lm}
    \bigl(1+a\sum_{\stackrel{\scriptstyle k=1}{k\neq l}}^{N}K_{lk}\bigr)
    -a(1-\delta_{lm})K_{lm},\label{eqn:Commutator_harmonic_Dunkl_1}\\
    & & [d_{l},d_{m}] = a(d_{m}-d_{l})K_{lm},
    \label{eqn:Commutator_harmonic_Dunkl_2}\\
    & & \alpha_{l} \cdot 1 = 0.
    \label{eqn:Action_harmonic_Dunkl}
  \end{eqnarray}
  \label{eqn:Needed_algebra_harmonic_Dunkl}
\end{subequations}
As we have mentioned, the phase factor of the difference product
part of the
ground state wave function can be
arbitrary. This phase factor affects the definition of
the Dunkl operators and
coordinate exchange operators
with hat, i.e., ${\hat \alpha}_{l}$, ${\hat \dalpha}_{l}$, ${\hat d}_{l}$
and ${\hat K_{lk}}$.
We have to introduce a phase factor in the defining relations of the
coordinate exchange operators \myref{eqn:Exchange_operator} and the
commutation relations of the Dunkl operators
\myref{eqn:Needed_algebra_harmonic_Dunkl}.~\cite{Ujino_4}
This modification is naturally
introduced by the inverse of the similarity transformation of the Dunkl
operators \myref{eqn:Inverse_harmonic_similar_transformation}.
Using the above relations, we can check that
the eigenstates of the Hamiltonian $I_{1}$ are
given by ~\cite{Brink_1,Brink_2}
\begin{subequations}
  \begin{eqnarray}
    \varphi_{\lambda} & = &
    \sum_{
    \stackrel{\scriptstyle \sigma:\;\mbox{\scriptsize distinct}}
    {\mbox{\scriptsize permutation}}}
    (\dalpha_{\sigma(1)})^{\lambda_{1}}
    (\dalpha_{\sigma(2)})^{\lambda_{2}}\cdots
    (\dalpha_{\sigma(N)})^{\lambda_{N}}
    \cdot 1\nonumber\\
    & = & m_{\lambda}(\dalpha_{1},\dalpha_{2},\cdots,
    \dalpha_{N})\cdot 1,
    \label{eqn:Monomial_eigenstates}\\
    I_{1}\varphi_{\lambda} & = & 
    E_{1}(\lambda)\varphi_{\lambda},
    \label{eqn:Monomial_eigenvalue_1st}
  \end{eqnarray}
  \label{eqn:Monomial_creation_operators}
\end{subequations}
where $m_{\lambda}(x_{1},x_{2},\cdots,x_{N})$ is
a monomial symmetric polynomial.~\cite{Stanley_1,Macdonald_1}
%Note that the summation over $S_{N}$ is taken such that any
term in the
%summand appears only once.
In terms of the Dunkl operators,
we can express the commuting conserved operators
$I_{n}$ \myref{eqn:Calogero_conserved_operators}
and the operators $V_{p}^{m}$
\myref{eqn:Calogero_Weyl_ordered_operators} as
\begin{eqnarray}
  I_{n} & = & 
  \sum_{l=1}^{N}(d_{l})^{n}\Bigr|_{\rm Sym},\;\;\; n=1,2,\cdots,N,
  \label{eqn:Calogero_conserved_operators_Dunkl}\\
  V_{p}^{m} & = & \sum_{l=1}^{N}
  \bigl[(\alpha_{l})^{m}(\dalpha_{l})^{p}\bigr]_{\rm W}
  \Bigr|_{\rm Sym},
  \label{eqn:Calogero_Weyl_ordered_operators_Dunkl}
\end{eqnarray}
where the symbol $\Bigr|_{\rm Sym}$ means that the action of
the operator
is restricted to symmetric functions.~\cite{Ujino_4} Then the
power sum creation operators are expressed by
\begin{equation}
  V_{n}^{0} = \sum_{l=1}^{N}
  (\dalpha_{l})^{n}\Bigr|_{\rm Sym}
  = p_{n}(\dalpha_{1},\dalpha_{2},\cdots,\dalpha_{N})\Bigr|_{\rm Sym},
  \label{eqn:Power_sum_creation_operators_Dunkl}
\end{equation}
where $p_{n}(x_{1},x_{2},\cdots,x_{N})$ is the power sum symmetric
polynomial
of degree $n$.~\cite{Stanley_1,Macdonald_1}
This shows that two kinds of eigenstates \myref{eqn:Eigenstates_QISM}
and \myref{eqn:Monomial_creation_operators} are related by the
transformation
between the power sum symmetric polynomials
and the monomial symmetric polynomials.

Next, we consider the QISM and the Dunkl operator formalism
for the Sutherland
model. The Lax matrix for the Sutherland model is
\begin{equation}
  \Lcal_{jk} = 
  p_{j}\delta_{jk}+{\rm i}a(1-\delta_{jk})\cot(x_{j}-x_{k}).
  \label{eqn:Lcal_matrix}
\end{equation}
The above Lax matrix gives the Hamiltonian of the Sutherland model by
\begin{eqnarray}
  \Hstilde & = & \frac{1}{2}\tsum\Lcal^{2} 
  + \frac{1}{6}a^{2}N(N-1)(N+1)
  \nonumber\\
  & = & \frac{1}{2}\tsum\Lcal^{2} + \epsilon_{\rm g},
  \label{eqn:Sutherland_model_in_Lcal}
\end{eqnarray}
where $\epsilon_{\rm g}$ is the ground state energy.
As is similar to eq. \myref{eqn:Calogero_ground_state_equation},
the ground
state of the Sutherland model satisfies the following equations,
\begin{eqnarray}
  & & \sum_{k=1}^{N}\Lcal_{jk}\psigtilde = 0,
  \mbox{ for }j=1,2,\cdots,N
  \nonumber\\
  & & \Rightarrow\Hstilde\psigtilde = \epsilon_{\rm g}\psigtilde,
  \label{eqn:Sutherland_ground_state_equation}
\end{eqnarray}
because the first term of the r.h.s. of 
eq. \myref{eqn:Sutherland_model_in_Lcal}
is a nonnegative operator. The ground state is given by the 
trigonometric Jastraw wave function:
\begin{equation}
  \psigtilde = \prod_{1\leq j < k\leq N}|\sin(x_{j}-x_{k})|^{a}.
  \label{eqn:Trigono_Jastraw}
\end{equation}
The phase factor of the above trigonometric Jastraw wave function
can be
arbitrary. 
By the change of the variables,
\begin{equation}
  \exp{2{\rm i}x_{j}}=z_{j},\;\;\; j=1,2,\cdots,N,
  \label{eqn:Variable_change}
\end{equation}
the Hamiltonian of the Sutherland model \myref{eqn:Sutherland_model} is
transformed to
\begin{equation}
  \Hstilde = -2\Bigl(\sum_{j=1}^{N}(z_{j}p_{z_{j}})^{2}
  +(a^{2}-a)\sum_{\stackrel{\scriptstyle j,k=1}{j\neq k}}^{N}
  \frac{z_{j}z_{k}}{(z_{j}-z_{k})^{2}}\Bigr),
  \label{eqn:Transformed_Sutherland_model}
\end{equation}
where $p_{z_{j}}=-{\rm i}\frac{\partial}{\partial z_{j}}$.
The ground state wave function \myref{eqn:Trigono_Jastraw} is
transformed to
\begin{equation}
  \psigtilde = \prod_{1\leq j<k\leq N}|z_{j}-z_{k}|^{a}
  \prod_{j=1}^{N}z_{j}^{-\frac{1}{2}a(N-1)}.
  \label{eqn:Trigono_Jastraw_in_z}
\end{equation}
Here we do not mind the difference of the scalar factor of
the ground state
wave function.
The similarity transformation of the above Hamiltonian yields
\begin{eqnarray}
  & & \Cmae\Hs-\epsilon_{\rm g}\nonumber\\
  & & = \psigtilde^{-1}(\Hstilde-\epsilon_{\rm g})\psigtilde
  \nonumber\\
  & & = -2\sum_{j=1}^{N}(z_{j}p_{z_{j}})^{2} + {\rm i}a
  \sum_{\stackrel{\scriptstyle j,k=1}{j\neq k}}^{N}
  \frac{z_{j}+z_{k}}{z_{j}-z_{k}}(z_{j}p_{z_{j}}-z_{k}p_{z_{k}}).
  \nonumber\\
  \label{eqn:Projected_Sutherland}
\end{eqnarray}
The above projected Hamiltonian can be derived from the Lax
matrices \myref{eqn:LQ-check_pair}. We define the ``ground state''
for the 
$\Lcheck$-matrix \myref{eqn:L-check_matrix} by the solution of
the equations,
\begin{subequations}
  \begin{equation}
    \sum_{k=1}^{N}\Lcheck_{kj}\Delta^{a} = 0, \mbox{ for }
    j=1,2,\cdots,N,
    \label{eqn:Calogero-Moser_ground_state_equation}
  \end{equation}
  and their solution is
  \begin{equation}
    \Delta^{a} = \prod_{1\leq j<k \leq N}
    |x_{j}-x_{k}|^{a}.
    \label{eqn:Jastraw}
  \end{equation}
\end{subequations}
The phase of the above Jastraw wave function also can be arbitrary.
The effect to the Dunkl operators for the Sutherland model,
which is made explicit by the
similarity transformation, is also the same as that of
the Calogero model.
By the similarity transformation using the above Jastraw function,
we define
$L$ and $Q$ as
\begin{subequations}
  \begin{eqnarray}
    L & = & \Delta^{-a}\Lcheck\Delta^{a},
    \label{eqn:L_matrix}\\
    Q & = & \Delta^{-a}\Qcheck\Delta^{a} = \Qcheck.
    \label{eqn:Q_matrix}
  \end{eqnarray}
\end{subequations}
Then we get the projected Hamiltonian \myref{eqn:Projected_Sutherland},
whose variables are not $\{z_{j}\}$ but $\{x_{j}\}$ 
by
\begin{eqnarray}
  \Hs-\epsilon_{\rm g} & = & 2\tsum(QL)^{2}
  \nonumber\\
  & \Define & 2 {\cal I}_{2}.
  \label{eqn:Sutherland_in_LQ}
\end{eqnarray}
>From now on, we take ${\cal I}_{2}$ as the Hamiltonian of the 
Sutherland model. The Jack symmetric polynomials
$J_{\lambda}(\vecvar{x};1/a)$ are uniquely defined by
\begin{subequations}
  \begin{eqnarray}
    & & \mbox{\hspace*{-20pt}(eigenfunction)} \nonumber\\
    & & \Cmae{\cal I}_{2}J_{\lambda}(\vecvar{x};1/a) \nonumber\\
    & & = \sum_{k=1}^{N}
    \bigl(\lambda_{k}^{2}+a(N+1-2k)\lambda_{k}\bigr)
    J_{\lambda}(\vecvar{x};1/a),
    \label{eqn:Jack_eigenfunction}\\
    & & \mbox{\hspace*{-20pt}(triangularity)} \nonumber\\
    & & \Cmae
    J_{\lambda}(\vecvar{x};1/a) = 
    \sum_{\mu\ledo\lambda}v_{\lambda\mu}(a)
    m_{\mu}(\vecvar{x}),
    \label{eqn:Jack_triangularity}\\
    & & \mbox{\hspace*{-20pt}(normalization)} \nonumber\\
    & & \Cmae v_{\lambda\lambda}(a) = 1,
    \label{eqn:Jack_normalization}
  \end{eqnarray}
  \label{eqn:Jack_definition}
\end{subequations}
where $\vecvar{x}=(x_{1},x_{2},\cdots,x_{N})$ and
$\lambda$ and $\mu$ are the Young tableaux \myref{eqn:Young_tableau}.
The symbol $\ledo$ is the dominance ordering among the Young
tableaux:~\cite{Stanley_1,Macdonald_1}
\begin{equation}
  \mu\ledo\lambda\Leftrightarrow\sum_{k=1}^{N}
  \mu_{k}=\sum_{k=1}^{N}\lambda_{k}
  \mbox{ and }\sum_{k=1}^{l}\mu_{k}\leq\sum_{k=1}^{l}\lambda_{k}
  \mbox{ for all } l.
  \label{eqn:Dominance_ordering}
\end{equation}
Note that the dominance ordering is not a total ordering but a
partial
ordering.  A total ordering among Young tableaux
is given by the lexicographic ordering:
\begin{eqnarray}
  \lefteqn{\mu\leq\lambda}\nonumber\\
  & \Leftrightarrow & 
  \sum_{k=1}^{N}\mu_{k}=\sum_{k=1}^{N}\lambda_{k}
  \mbox{ and the first nonvanishing}\nonumber\\
  & &
  \mbox{difference }\lambda_{l}-\mu_{l}>0.
  \label{eqn:Lexicographic_ordering}
\end{eqnarray}
Commuting conserved operators of the Sutherland model are
given by
\begin{equation}
  {\cal I}_{n} = \tsum(QL)^{n}.
  \label{eqn:Sutherland_conserved_operator_QISM}
\end{equation}
We have similar relations to the generalized Lax equations 
for the Calogero model \myref{eqn:Calogero_generalized_Lax},
\begin{subequations}
  \begin{eqnarray}
    \bigl[U_{q}^{l},L\bigr] & = &
    \bigl[L,Y_{q}^{l}\bigr]-q[(L)^{l}(Q)^{q-1}]_{\rm W},
    \label{eqn:Sutherland_generalized_Lax_L}\\
    \bigl[U_{q}^{l},Q\bigr] & = & \bigl[Q,Y_{q}^{l}\bigr]
    +m[(L)^{l}(Q)^{q}]_{\rm W},
    \label{eqn:Sutherland_generalized_Lax_Q}
  \end{eqnarray}
  \label{eqn:Sutherland_generalized_Lax}
\end{subequations}
where the operator $U_{q}^{l}$ is defined by
\begin{equation}
  U_{q}^{l} = \tsum[(L)^{l}(Q)^{q}]_{\rm W}.
  \label{eqn:Sutherland_Weyl_ordered_operators}
\end{equation}
The operator-valued matrix $Y_{q}^{l}$ also satisfies the
sum-to-zero condition,
\begin{equation}
  \sum_{j=1}^{N}(Y_{q}^{l})_{jk} = 
  \sum_{j=1}^{N}(Y_{q}^{l})_{kj} = 0,\;\;\;
  \mbox{for }k=1,2,\cdots,N.
  \label{eqn:Sutherland_sum-to-zero}
\end{equation}
Since eqs. \myref{eqn:Calogero_generalized_Lax} and
\myref{eqn:Sutherland_generalized_Lax} has the same form,
we notice the
correspondence between the Calogero model and the Sutherland model:
\begin{equation}
  \Lm = L,
  \Lp\leftrightarrow Q,
  I_{k}\leftrightarrow {\cal I}_{k}.
  \label{eqn:Correspondence_two_QISM}
\end{equation}
This means that the two QISM's for the Calogero and
Sutherland models give
two different representations of the same commutator algebra.

The same situation can also be observed in the Dunkl
operators' approach.
We introduce the Dunkl
operators for the Sutherland model whose action on
the ground state is removed
in a similar way to deal with the Dunkl operators
for the Calogero model:
\begin{subequations}
  \begin{eqnarray}
    \nabla_{l} & = & {\rm i}\bigl(p_{l}
    +{\rm i}a\sum_{\stackrel{\scriptstyle k=1}{k\neq l}}^{N}
    \frac{1}{x_{l}-x_{k}}(K_{lk}-1)\bigr),
    \label{eqn:Coupled_momentum}\\
    x_{l}, & &
    \label{eqn:coordinate}\\
    D_{l} & = & x_{l}\nabla_{l}.
    \label{eqn:D_operator}
  \end{eqnarray}
  \label{eqn:Sutherland_Dunkl}
\end{subequations}
These Dunkl operators satisfies the following relations,
\begin{subequations}
  \begin{eqnarray}
    & & [\nabla_{l},\nabla_{m}] = 0,\;\;\;
    [x_{l},x_{m}] = 0,
    \label{eqn:Commutator_Sutherland_Dunkl_0}\\
    & & [\nabla_{l},x_{m}]
    \nonumber\\
    & & \;\;= \delta_{lm}
    \bigl(1+a\sum_{\stackrel{\scriptstyle k=1}{k\neq l}}^{N}K_{lk}\bigr)
    -a(1-\delta_{lm})K_{lm},\label{eqn:Commutator_Dunkl_Sutherland_1}\\
    & & [D_{l},D_{m}] = a(D_{m}-D_{l})K_{lm},
    \label{eqn:Commutator_Dunkl_Sutherland_2}\\
    & & \nabla_{l} \cdot 1 = 0,
    \label{eqn:Action_Dunkl_Sutherland}
  \end{eqnarray}
  \label{eqn:Needed_algebra_Dunkl_Sutherland}
\end{subequations}
which are completely the same as those of Dunkl operators for
the Calogero model \myref{eqn:Needed_algebra_harmonic_Dunkl}.
Commuting conserved operators
\myref{eqn:Sutherland_conserved_operator_QISM}
are written by the Dunkl operator as
\begin{equation}
  {\cal I}_{n}=\sum_{l=1}^{N}(D_{l})^{n}\Bigr|_{\rm Sym},
  \;\;n=1,2,\cdots,N.
  \label{eqn:Sutherland_conserved_operators_Dunkl}
\end{equation}
Thus we notice the correspondence between the two sets
of Dunkl operators:
\begin{equation}
  \alpha_{l} = \nabla_{l},
  \dalpha_{l} \leftrightarrow x_{l},
  d_{l} \leftrightarrow D_{l}.
  \label{eqn:Correspondence_two_Dunkl}
\end{equation}
Moreover,
in the limit $\omega\rightarrow\infty$, the Lax matrices
and the Dunkl
operators for the Calogero model reduce to those of
the Sutherland model.
Thus our theory for the Hi-Jack symmetric polynomials
described by
the Dunkl operators for the Calogero model contains
the results for
the Jack symmetric polynomials written by the Dunkl
operators for the Sutherland model.

We have summarized the QISM and the Dunkl operators' approach
to the Calogero and Sutherland models. They give
two different representations of the same commutator algebra.
The QISM and the Dunkl operators for the Calogero model
include those for the Sutherland model as a special case
$\omega\rightarrow\infty$. Thus we can say that the theory of
the Calogero
model and the Hi-Jack polynomials
is a one-parameter deformation of that of the Sutherland
model and the Jack
polynomials.
In the following section, we investigate the Hi-Jack
polynomials using the
Dunkl operators.

\section{Hi-Jack Symmetric Polynomials}\label{sec:Hi-Jack}
Following the definition of the Jack symmetric polynomials
\myref{eqn:Jack_definition},
we define the Hi-Jack symmetric polynomials
$j_{\lambda}(\vecvar{x};\omega,1/a)$ by
%\begin{definition}
  \begin{subequations}
    \begin{eqnarray}
      & & \mbox{\hspace*{-20pt} (eigenfunction)} \nonumber\\
      & & \Cmae
      I_{1}j_{\lambda}(\vecvar{x};\omega,1/a)=\sum_{k=1}^{N}
      \lambda_{k}
      j_{\lambda}(\vecvar{x};\omega,1/a)\nonumber\\
      & & = E_{1}(\lambda)j_{\lambda}(\vecvar{x};\omega,1/a),
      \label{eqn:Hi-Jack_eigenfunction_1}\\
      & & \Cmae I_{2}j_{\lambda}(\vecvar{x};\omega,1/a)
      \nonumber\\
      & & = \sum_{k=1}^{N}
      \bigl(\lambda_{k}^{2}+a(N+1-2k)\lambda_{k}\bigr)
      j_{\lambda}(\vecvar{x};\omega,1/a)\nonumber\\
      & & = E_{2}(\lambda)j_{\lambda}(\vecvar{x};\omega,1/a),
      \label{eqn:Hi-Jack_eigenfunction_2}\\
      & & \mbox{\hspace*{-20pt}(triangularity)} \nonumber\\
      & & \Cmae j_{\lambda}(\vecvar{x};\omega,1/a)
      = \sum_{\stackrel{\scriptstyle \mu\ledo\lambda}
      {\mbox{\scriptsize or }|\mu|<|\lambda|}}
      w_{\lambda\mu}(a,1/2\omega)
      m_{\mu}(\vecvar{x}),
      \label{eqn:Hi-Jack_triangularity}\\
      & & \mbox{\hspace*{-20pt}(normalization)} \nonumber\\
      & & \Cmae w_{\lambda\lambda}(a,1/2\omega) = 1,
      \label{eqn:Hi-Jack_normalization}
    \end{eqnarray}
    \label{eqn:Hi-Jack_definition}
  \end{subequations}
%\end{definition}
where $|\lambda|$ is the weight of the Young tableau,
$|\lambda|=\sum_{k=1}^{N}\lambda_{k}$.
In order to write down the Rodrigues formula for
the Hi-Jack polynomials,
it is convenient to introduce the following operators
(cf. eq. \myref{eqn:Dunkl}):
\begin{subequations}
  \begin{eqnarray}
    \dalpha_{J} & = & \prod_{j\in J}\dalpha_{j},
    \label{eqn:Monomial_generator}\\
    d_{m,J} & = & (d_{j_{1}} + ma)(d_{j_{2}} + (m+1)a) 
    \nonumber \\
    & & \cdots(d_{j_{k}} + (m+k-1)a),
    \label{eqn:Q-det}
  \end{eqnarray}
  \label{eqn:Parts_of_generator}
\end{subequations}
where $J$ is a subset of a set $\{1,2,\cdots,N\}$
whose number of elements $|J|$ is
equal to $k$, $J\subseteq\{1,2,\cdots,N\}$, $|J|=k$. 
>From eq. \myref{eqn:Commutator_harmonic_Dunkl_2},
we can verify an identity,
\begin{eqnarray}
  & & \mbox{\hspace*{-20pt}}(d_{i}+ma)(d_{j}+(m+1)a)\Sym^{\{i,j\}}
  \nonumber\\
  & & = (d_{j}+ma)(d_{i}+(m+1)a)\Sym^{\{i,j\}},
  \label{eqn:Identity_di_and_dj}
\end{eqnarray}
where $m$ is some integer. The symbol $\Sym^{J}$ where $J$ is
some set of
integers means that the action of operators is restricted to
the space that
is symmetric with respect to exchange of any indices in the set $J$.
This
identity guarantees that the operator $d_{m,J}$ does not depend
on the ordering
of the elements of a set $J$.
The generators of the Hi-Jack polynomials are expressed as
\begin{subequations}
  \begin{eqnarray}
    & & \Cmae\Cmae b^{+}_{k} =
    \sum_{\stackrel{\scriptstyle J\subseteq
    \{1,2,\cdots,N\}}{|J|=k}}
    \dalpha_{J}d_{1,J},\;\mbox{for }
    k=1,2,\cdots,N-1,
    \label{eqn:Generator_bulk}\\
    & & \Cmae\Cmae b^{+}_{N} =
    \dalpha_{1}\dalpha_{2}\cdots\dalpha_{N}.
    \label{eqn:Generator_edge}
  \end{eqnarray}
  \label{eqn:Generator}
\end{subequations}
Using the
generators \myref{eqn:Generator}, we can write down
the Rodrigues formula
for the Hi-Jack polynomials $j_{\lambda}(\vecvar{x};\omega,1/a)$
as
\begin{eqnarray}
  j_{\lambda}(\vecvar{x};\omega,1/a) & = & 
  C_{\lambda}^{-1}(b_{N}^{+})^{\lambda_{N}}
  (b_{N-1}^{+})^{\lambda_{N-1}-\lambda_{N}}\nonumber \\
  & & \cdots (b_{1}^{+})^{\lambda_{1}-\lambda_{2}}\cdot 1,
  \label{eqn:Hi-Jack_polynomials}
\end{eqnarray}
with the constant $C_{\lambda}$ given by
\begin{equation}
  C_{\lambda}=
  \prod_{k=1}^{N-1}C_{k}
  (\lambda_{1},\lambda_{2},\cdots,\lambda_{k+1};a),
  \label{eqn:Normalization}
\end{equation}
where
\begin{eqnarray}
  & & \Mae C_{k}(\lambda_{1},\lambda_{2},\cdots,\lambda_{k+1};a)
  \nonumber\\
  & & = (a)_{\lambda_{k}-\lambda_{k+1}}
  (2a+\lambda_{k-1}-\lambda_{k})_{\lambda_{k}-\lambda_{k+1}}
  \nonumber \\
  & & \Cusr\cdots
  (ka+\lambda_{1}-\lambda_{k})_{\lambda_{k}-\lambda_{k+1}}.
  \label{eqn:Parts_of_normalization}
\end{eqnarray}
In the above expression, the symbol $(\beta)_{n}$ is the
Pochhammer symbol, that is,
$(\beta)_{n}=\beta(\beta+1)\cdots(\beta+n-1)$,
$(\beta)_{0}\Define 1$.
What we want to prove is summarized as the following proposition.
\begin{proposition}
  The symmetric polynomials generated by
  the Rodrigues formula \myref{eqn:Hi-Jack_polynomials} satisfy
  the definition of
  the Hi-Jack symmetric polynomials \myref{eqn:Hi-Jack_definition}.
  \label{prop:Hi-Jack}
\end{proposition}

The first two out of four requirements \myref{eqn:Hi-Jack_definition}
are derived from the following propositions.
\begin{proposition}
  \begin{equation}
    \bigl[I_{1},b_{k}^{+}\bigr]\Bigr|_{\rm Sym} =
    k b_{k}^{+}\Bigr|_{\rm Sym}.
    \label{eqn:Commutator_I1_bk+}
  \end{equation}
  \label{prop:Homogeneous}
\end{proposition}
\begin{proposition}
  The null operators $n_{i+1,J}$, which are defined by
  \begin{equation}
    n_{k+1,J} = d_{0,J},\;J\subseteq\{1,2,\cdots,N\},\;|J|=k+1,
    \label{eqn:Null_operator}
  \end{equation}
  satisfy
  \begin{equation}
    n_{k+1,J} (b_{k}^{+})^{\lambda_{k}}
    (b_{k-1}^{+})^{\lambda_{k-1}-\lambda_{k}} \cdots
    (b_{1}^{+})^{\lambda_{1}-\lambda_{2}}\cdot 1 = 0.
    \label{eqn:Null_vanishes}
  \end{equation}
  \label{prop:Null_operator}
\end{proposition}
\begin{proposition}
  \begin{eqnarray}
    \Cmae\Cmae\bigl[I_{2},\bp_{k}\bigr]\Sym
    & = & \Bigl\{b_{k}^{+}\bigl(2I_{1}+k+ak(N-k)\bigr)\nonumber\\
    \Cmae\Cmae & & +
    \sum_{\stackrel{\scriptstyle J\subseteq\{1,2,\cdots,N\}}{|J|=k+1}}
    g_{k+1,J}n_{k+1,J}\Bigr\}\Sym,
    \label{eqn:Eigenvalue_precursor}
  \end{eqnarray}
  where $g_{k+1,J}$ is an unspecified nonsingular operator
  that satisfies
   $g_{N+1,J}=0$.
  \label{prop:Eigenvalue_precursor}
\end{proposition}

The first requirement directly follows from Proposition
\ref{prop:Homogeneous}.
For a while, we forget about the normalization constant.
>From the l.h.s.
of eq. \myref{eqn:Hi-Jack_eigenfunction_1}, we get
\begin{eqnarray}
  & & \Cmae I_{1}(b_{N}^{+})^{\lambda_{N}}
  (b_{N-1}^{+})^{\lambda_{N-1}-\lambda_{N}} \cdots
  (b_{1}^{+})^{\lambda_{1}-\lambda_{2}}\cdot 1\nonumber\\
  & & = \Bigl(\bigl[I_{1},(b_{N}^{+})^{\lambda_{N}}
  (b_{N-1}^{+})^{\lambda_{N-1}-\lambda_{N}} \cdots
  (b_{1}^{+})^{\lambda_{1}-\lambda_{2}}\bigr]\nonumber\\
  & & \Cusr +(b_{N}^{+})^{\lambda_{N}}
  (b_{N-1}^{+})^{\lambda_{N-1}-\lambda_{N}} \cdots
  (b_{1}^{+})^{\lambda_{1}-\lambda_{2}}I_{1}\Bigr)\cdot 1.
  \nonumber\\
  \label{eqn:Requirement1-1}
\end{eqnarray}
Because of eq. \myref{eqn:Action_harmonic_Dunkl},
the second term of the
above equation vanishes:
\begin{equation}
  I_{1}\cdot 1 = \sum_{k=1}^{N} \dalpha_{k}\alpha_{k}\cdot 1 = 0.
  \label{eqn:Requirement1-2}
\end{equation}
Then using Proposition \ref{prop:Homogeneous},
we get the expected result:
\begin{eqnarray}
  & & \Cmae I_{1}(b_{N}^{+})^{\lambda_{N}}
  (b_{N-1}^{+})^{\lambda_{N-1}-\lambda_{N}} \cdots
  (b_{1}^{+})^{\lambda_{1}-\lambda_{2}}\cdot 1\nonumber\\
  & & = \Bigl(\sum_{k=1}^{N-1}k(\lambda_{k}-\lambda_{k+1})
  + N\lambda_{N}\Bigr)
  \nonumber\\
  & & \Cusr (b_{N}^{+})^{\lambda_{N}}
  (b_{N-1}^{+})^{\lambda_{N-1}-\lambda_{N}} \cdots
  (b_{1}^{+})^{\lambda_{1}-\lambda_{2}}\cdot 1
  \nonumber\\
  & & = \sum_{k=1}^{N}\lambda_{k}
  (b_{N}^{+})^{\lambda_{N}}
  (b_{N-1}^{+})^{\lambda_{N-1}-\lambda_{N}} \cdots
  (b_{1}^{+})^{\lambda_{1}-\lambda_{2}}\cdot 1.\nonumber\\
  \label{eqn:Requirement1-3}
\end{eqnarray}
The second requirement \myref{eqn:Hi-Jack_eigenfunction_2}
is shown by induction. It is easy to show it for $\lambda=0$,
\begin{equation}
  I_{2}\cdot j_{0}(\vecvar{x};\omega,1/a)
  =E_{2}(0)j_{0}(\vecvar{x};\omega,1/a)=0,
  \label{eqn:Requirement2-1}
\end{equation}
by using eq. \myref{eqn:Action_harmonic_Dunkl}
because $j_{0}(\vecvar{x};\omega,1/a)$ is equal to $1$
as a polynomial.
By inductive assumption, eq. \myref{eqn:Hi-Jack_eigenfunction_2}
holds up to $\lambda=\{\lambda_{1},\lambda_{2},\cdots,\lambda_{k},
\underbrace{0,\cdots,0}_{N-k}\}$. Then for 
$\lambda=\{\lambda_{1}+1,\lambda_{2}+1,\cdots,\lambda_{k}+1,
\underbrace{0,\cdots,0}_{N-k}\}$, we have
\begin{eqnarray}
  & & \Mae I_{2}b_{k}^{+}(b_{k}^{+})^{\lambda_{k}}
  (b_{k-1}^{+})^{\lambda_{k-1}-\lambda_{k}}\cdots
  (b_{1}^{+})^{\lambda_{1}-\lambda_{2}}\cdot 1 \nonumber\\
  & & = \Bigl(\bigl[I_{2},b_{k}^{+}\bigr]+b_{k}^{+}I_{2}\Bigr)
  \nonumber\\
  & & \Cusr (b_{k}^{+})^{\lambda_{k}}
  (b_{k-1}^{+})^{\lambda_{k-1}-\lambda_{k}}\cdots
  (b_{1}^{+})^{\lambda_{1}-\lambda_{2}}\cdot 1.
  \label{eqn:Requirement2-2}
\end{eqnarray}
>From the inductive assumption and Proposition
\ref{prop:Eigenvalue_precursor},
we get
\begin{eqnarray}
  & & \Mae I_{2}b_{k}^{+}(b_{k}^{+})^{\lambda_{k}}
  (b_{k-1}^{+})^{\lambda_{k-1}-\lambda_{k}}\cdot
  (b_{1}^{+})^{\lambda_{1}-\lambda_{2}}\cdot 1\nonumber\\
  & & = \Bigl(E_{2}
  (\{\lambda_{1},\lambda_{2},\cdots,\lambda_{k},0,\cdots,0\})
  \nonumber\\
  & & \Cusr
  +2\sum_{j=1}^{k}\lambda_{j} + k + ak(N-k)\Bigr)\nonumber\\
  & & \Cusr\times (b_{k}^{+})^{\lambda_{k}}
  (b_{k-1}^{+})^{\lambda_{k-1}-\lambda_{k}}\cdots
  (b_{1}^{+})^{\lambda_{1}-\lambda_{2}}\cdot 1\nonumber\\
  & & =  E_{2}
  (\{\lambda_{1}+1,\lambda_{2}+1,\cdots,\lambda_{k}+1,0,\cdots,0\})
  \nonumber\\
  & & \Cusr\times (b_{k}^{+})^{\lambda_{k}}
  (b_{k-1}^{+})^{\lambda_{k-1}-\lambda_{k}}\cdots
  (b_{1}^{+})^{\lambda_{1}-\lambda_{2}}\cdot 1,
  \label{eqn:Requirement2-3}
\end{eqnarray}
which completes the proof.

As byproducts of the proof of the last two requirements of Proposition
\ref{prop:Hi-Jack}, we notice the following results.
\begin{proposition}
  The expansion coefficients of the Hi-Jack polynomials
  $C_{\lambda}w_{\lambda\mu}(a,1/2\omega)$ are
  polynomials of $a$ and $1/2\omega$ with integer coefficients.
  This property is analogous to that stated by
  the Macdonald-Stanley conjecture for the Jack
  polynomials.~\cite{Stanley_1,Macdonald_1,Lapointe_2}
  \label{prop:Macdonald-Stanley_Hi-Jack}
\end{proposition}
\begin{proposition}
  \begin{equation}
    J_{\lambda}(\dalpha_{1},\dalpha_{2},\cdots,\dalpha_{N};1/a)\cdot 1
    = j_{\lambda}(\vecvar{x};\omega,1/a).
    \label{eqn:Jack_to_Hi-Jack}
  \end{equation}
  \label{prop:Jack_to_Hi-Jack}
\end{proposition}
During the discussion in
\S\ref{sec:preparations},
we have noticed that the Hi-Jack polynomials should
reduce to the Jack polynomials in the limit $\omega\rightarrow\infty$:
\begin{equation}
  j_{\lambda}(\vecvar{x};\omega=\infty,1/a)=J_{\lambda}(\vecvar{x};1/a).
  \label{eqn:Hi-Jack_to_Jack}
\end{equation}
Proposition \ref{prop:Jack_to_Hi-Jack} gives
another relationship between
the Jack polynomials and the Hi-Jack polynomials.
The relationship between the eigenstates of the Hamiltonian $\Hc$
given by the QISM \myref{eqn:Eigenstates_QISM} or the eigenstates
\myref{eqn:Monomial_eigenstates} and the Hi-Jack polynomials is now
clear. Several bases for the ring of homogeneous symmetric
polynomials are known.~\cite{Stanley_1,Macdonald_1} 
The power sums, the
monomial symmetric polynomials and the Jack polynomials
are examples of such
bases. Thus, the transformation between the Hi-Jack
polynomials and two kinds
of the eigenstates of the Hamiltonian $\Hc$ is
the transformation
between the bases of homogeneous symmetric polynomials.
Defining the transformations by
\begin{subequations}
  \begin{eqnarray}
    f^{pJ}_{\lambda}(\{p_{\lambda}\})=J_{\lambda},
    \label{eqn:Power-sum_to_Jack}\\
    f^{mJ}_{\lambda}(\{m_{\lambda}\})=J_{\lambda},
    \label{eqn:Monomial_to_Jack}
  \end{eqnarray}
\end{subequations}
we have 
\begin{subequations}
  \begin{eqnarray}
    f^{pJ}_{\lambda}(\{\phi_{\lambda}\})=j_{\lambda},\\
    f^{mJ}_{\lambda}(\{\varphi_{\lambda}\})=j_{\lambda}.
  \end{eqnarray}
  \label{eqn:Basis_transformation}
\end{subequations}
Note that the transformation \myref{eqn:Monomial_to_Jack}
is nothing but the
expansion of the Jack polynomials in the monomial symmetric
polynomials
\myref{eqn:Jack_triangularity}.

In the next section, we shall prove Proposition
\ref{prop:Homogeneous} -- 
Proposition \ref{prop:Jack_to_Hi-Jack}
and the last two requirements of Proposition \ref{prop:Hi-Jack}.

\section{Proofs}\label{sec:proofs}

\subsection{Hamiltonian}\label{subsec:hamiltonian}
We shall prove Proposition \ref{prop:Homogeneous}. It is easy to prove
the case $k=N$. From the definition of $I_{1}$ and $b_{N}^{+}$, we have
\begin{eqnarray}
  & & \Mae\bigl[I_{1},b_{N}^{+}\bigr]\nonumber\\
  & & = \sum_{j=1}^{N}\bigl[d_{j},\dalpha_{1}\dalpha_{2}\cdots\dalpha_{N}
  \bigr]\nonumber\\
  & & = \sum_{j=1}^{N}\sum_{l=1}^{N}\dalpha_{1}\cdots\dalpha_{l-1}
  \bigl[d_{j},\dalpha_{l}\bigr]\dalpha_{l+1}\cdots\dalpha_{N}.
  \label{eqn:k=N_1}
\end{eqnarray}
Using eq. \myref{eqn:Commutator_harmonic_Dunkl_2}, we get
\begin{eqnarray}
  \bigl[I_{1},b_{N}^{+}\bigr] & = & 
  \sum_{j=1}^{N}\sum_{l=1}^{N}\dalpha_{1}\cdots\dalpha_{l-1}
  \nonumber\\
  & & \dalpha_{j}\bigl(\delta_{jl}(1+a\sum_{\stackrel{\scriptstyle i=1}
  {i\neq j}}^{N}K_{ji})-a(1-\delta_{jl})K_{jl}\bigr)
  \nonumber\\
  & & \dalpha_{l+1}\cdots\dalpha_{N}\nonumber\\
  & = & \bigl(Nb_{N}^{+} + a\dalpha_{1}\cdots\dalpha_{l-1}
  \nonumber\\
  & & \sum_{\stackrel{\scriptstyle j,l=1}{j\neq l}}^{N}
  \dalpha_{j}(K_{jl}-K_{jl})\dalpha_{l+1}\cdots\dalpha_{N}\bigr)
  \nonumber\\
  & = & Nb_{N}^{+},
  \label{eqn:k=N_2}
\end{eqnarray}
which says the validity of eq. \myref{eqn:Commutator_I1_bk+}
for the case $k=N$. Note that we do not have to restrict
the action of the operator in the above calculation.

For the case $k\neq N$, we need more computation. First,
we decompose the
Hamiltonian $I_{1}$ into two parts:
\begin{equation}
  \bigl[I_{1},b_{k}^{+}\bigr]
  = \sum_{\stackrel{\scriptstyle J\subseteq\{1,2,\cdots,N\}}{|J|=k}}
  \bigl[\sum_{i\in J}d_{i}+
  \sum_{i\notin J}d_{i},\dalpha_{J}d_{1,J}\bigr].
  \label{eqn:k=generic_1}
\end{equation}
The first part of the r.h.s. of eq. \myref{eqn:k=generic_1}
is calculated as
\begin{eqnarray}
  & & \!\!\bigl[\sum_{i\in J}d_{i},\dalpha_{J}d_{1,J}\bigr]
  \nonumber\\
  & & = \sum_{i\in J}\sum_{l=1}^{k}\Bigl\{\dalpha_{j_{1}}\cdots
  \dalpha_{j_{l-1}}\bigl[d_{i},\dalpha_{j_{l}}\bigr]
  \dalpha_{j_{l+1}}\cdots\dalpha_{l_{k}}d_{1,J}\nonumber\\
  & & \Cusr + \dalpha_{J}(d_{j_{1}}+a)\cdots(d_{j_{l-1}}+(l-1)a)
  \bigl[d_{i},d_{j_{l}}+la\bigr]\nonumber\\
  & & \Cusr\Cusr (d_{j_{l+1}}+(l+1)a)\cdots(d_{l_{k}}+ka)\Bigr\}
  \nonumber\\
  & & = \sum_{i\in J}\sum_{l=1}^{k}\Bigl\{\dalpha_{j_{1}}\cdots
  \dalpha_{j_{l-1}}\nonumber\\
  & & \Cusr\bigl(\delta_{ij_{l}}\dalpha_{j_{l}}
  (1+a\sum_{\stackrel{\scriptstyle j\in J}{j\neq j_{l}}}K_{j_{l}j}
  +a\sum_{j\notin J}K_{j_{l}j})
  -(1-\delta_{ij_{l}})aK_{ij_{l}}\bigr)\nonumber\\
  & & \Usr\dalpha_{j_{l+1}}\cdots\dalpha_{l_{k}}d_{1,J}\nonumber\\
  & & \Cusr + \dalpha_{J}(d_{j_{1}}+a)\cdots(d_{j_{l-1}}+(l-1)a)
  \nonumber\\
  & & \Usr\bigl((d_{j_{l}}+la)-(d_{i}+la)\bigr)K_{ij_{l}}
  \nonumber\\
  & & \Usr (d_{j_{l+1}}+(l+1)a)\cdots(d_{l_{k}}+ka)\Bigr\}.
  \label{eqn:k=generic_2}
\end{eqnarray}
We move exchange operators to the rightmost and utilize the 
restriction $\Sym$.
Using eqs. \myref{eqn:Exchange_operator}
and \myref{eqn:Identity_di_and_dj}, we get
\begin{eqnarray}
  & & \!\!
  \bigl[\sum_{i\in J}d_{i},\dalpha_{J}d_{1,J}\bigr]\Sym\nonumber\\
  & & = k\alpha_{J}d_{1,J}\Sym + a\sum_{j\notin J}
  \sum_{l=1}^{k}\dalpha_{J}
  d_{1,J\setminus\{j_{l}}\}(d_{i}+ka)\Sym\nonumber\\
  & & \Cusr +\frac{1}{2}a\sum_{l=1}^{k}
  \sum_{\stackrel{\scriptstyle i\in J}{i\neq j_{l}}}
  \Bigl\{\bigl(\dalpha_{J\setminus\{i\}}\dalpha_{j_{l}}
  -\dalpha_{J\setminus\{j_{l}\}}\dalpha_{i}\bigr)d_{1,J}\nonumber\\
  & & \Cusr +\dalpha_{J}\bigl(d_{1,J\setminus\{i\}}(d_{j_{l}}+ka)
  -d_{1,J\setminus\{j_{l}\}}(d_{i}+ka)\bigr)\Bigr\}\Sym\nonumber\\
  & & = k\alpha_{J}d_{1,J}\Sym + a\sum_{i\notin J}\sum_{l=1}^{k}
  \dalpha_{J}
  d_{1,J\setminus\{j_{l}\}}(d_{i}+ka)\Sym.\nonumber\\
  \label{eqn:k=generic_3}
\end{eqnarray}
The second part of the r.h.s. of eq. \myref{eqn:k=generic_1} is
calculated as
\begin{eqnarray}
  & & \Cmae\bigl[\sum_{i\notin J} d_{i},\dalpha_{J}d_{1,J}\bigr]\Sym
  \nonumber\\
  & & = \sum_{i\notin J}\sum_{l=1}^{k}\Bigl\{\dalpha_{j_{1}}\cdots
  \dalpha_{j_{l-1}}\bigl[d_{i},\dalpha_{j_{l}}\bigr]\dalpha_{j_{l+1}}
  \cdots\dalpha_{j_{k}}d_{1,J}\nonumber\\
  & & \Cusr + \dalpha_{J}(d_{j_{1}}+a)\cdots(d_{j_{l-1}}+(l-1)a)
  \bigl[d_{i},d_{j_{l}}+la\bigr]\nonumber\\
  & & \Cusr\Cusr (d_{j_{l+1}}+(l+1)a)\cdots(d_{l_{k}}+ka)\Bigr\}\Sym
  \nonumber\\
  & & = \sum_{i\notin J}\sum_{l=1}^{k}\Bigl\{-a\dalpha_{j_{1}}\cdots
  \dalpha_{j_{l-1}}\dalpha_{i}K_{ij_{l}}\dalpha_{j_{l+1}}
  \cdots\dalpha_{j_{k}}d_{1,J}\nonumber\\
  & & \Cusr + \dalpha_{J}(d_{j_{1}}+a)\cdots(d_{j_{l-1}}+(l-1)a)
  \nonumber\\
  & & \Cusr\Cusr
  \bigl((d_{j_{l}}+la) - (d_{j_{i}}+la)\bigr)K_{ij_{l}}\nonumber\\
  & & \Cusr\Cusr(d_{j_{l+1}}+(l+1)a)\cdots(d_{l_{k}}+ka)\Bigr\}\Sym
  \nonumber\\
  & & = \sum_{i\notin J}\sum_{l=1}^{k}
  \Bigl\{-a\dalpha_{J\setminus\{j_{l}\}}
  \dalpha_{i}d_{1,J\setminus\{j_{l}\}}(d_{i}+ka)\nonumber\\
  & & \Cusr 
  + \dalpha_{J}\bigl(d_{1,J}-d_{1,J\setminus\{j_{l}\}}(d_{i}+ka)\bigr)
  \Bigr\}\Sym.
  \label{eqn:k=generic_4}
\end{eqnarray}
Substitution of eqs. \myref{eqn:k=generic_3} and
\myref{eqn:k=generic_4}
into eq. \myref{eqn:k=generic_1} yields
\begin{eqnarray}
  & & \Cmae\bigl[I_{1},b_{k}^{+}\bigr]\Sym\nonumber\\
  & & = 
  \sum_{\stackrel{\scriptstyle J\subseteq\{1,2,\cdots,N\}}{|J|=k}}
  \Bigl\{k\alpha_{J}d_{1,J}\nonumber\\
  & &  \Cusr + a\sum_{i\notin J}\sum_{l=1}^{k}\bigl\{
  (\dalpha_{J}-\dalpha_{J\setminus\{j_{l}\}}\dalpha_{i})
  d_{1,J\setminus\{j_{l}\}}(d_{i}+ka)\nonumber\\
  & & \Cusr 
  + \dalpha_{J}\bigl(d_{1,J}-d_{1,J\setminus\{j_{l}\}}(d_{i}+ka)\bigr)
  \bigr\}\Bigr\}\Sym\nonumber\\
  & & = kb_{k}^{+}\Sym.
  \label{eqn:k=generic_5}
\end{eqnarray}
Thus we have proved Proposition \ref{prop:Homogeneous}.

\subsection{Null operators}\label{subsec:null_operators}
Since the function
$(b_{k}^{+})^{\lambda_{k}}(b_{k-1}^{+})^{\lambda_{k-1}-\lambda_{k}}
\cdots
(b_{1}^{+})^{\lambda_{1}-\lambda_{2}}\cdot 1$ is a symmetric
function of
$\{x_{1},x_{2},\cdots,x_{N}\}$, it is sufficient to
prove the case
$J=\{1,2,\cdots,k+1\}$. For brevity, we use the symbol
$n_{k+1}=n_{k+1,\{1,\cdots,k+1\}}$ hereafter.
Then the expression to be proved is
\begin{equation}
  n_{k+1}(b_{k}^{+})^{\lambda_{k}}
  (b_{k-1}^{+})^{\lambda_{k-1}-\lambda_{k}}
  \cdots(b_{1}^{+})^{\lambda_{1}-\lambda_{2}}\cdot 1=0.
  \label{eqn:Null_1}
\end{equation}
This follows from
\begin{equation}
  \bigl[n_{i+1},\bp_{k}\bigr]\Sym\sim 
  n_{k+1}\Sym, \mbox{ for }i\geq k,
  \label{eqn:Null_2}
\end{equation}
where the symbol $\sim$ means that the term on the r.h.s. can 
be multiplied
on the left by some nonsingular operator ${\cal O}$, 
${\cal O}\cdot 0 = 0$.
We can easily verify
\begin{eqnarray}
  & & \Mae n_{k+1}(b_{k}^{+})^{\lambda_{k}}
  (b_{k-1}^{+})^{\lambda_{k-1}-\lambda_{k}}
  \cdots(b_{1}^{+})^{\lambda_{1}-\lambda_{2}}\cdot 1\nonumber\\
  & & \sim n_{k+1}(b_{k-1}^{+})^{\lambda_{k-1}-\lambda_{k}}
  \cdots(b_{1}^{+})^{\lambda_{1}-\lambda_{2}}\cdot 1\nonumber\\
  & \vdots & \nonumber\\
  & & \sim n_{2}\cdot 1\nonumber\\
  & & = 0,
  \label{eqn:Null_3}
\end{eqnarray}
using eqs. \myref{eqn:Action_harmonic_Dunkl} and \myref{eqn:Null_2}.
For convenience of explanation, we introduce a symbol $[m]$ for a set
$\{1,2,\cdots,m\}$ with some integer $m$. We also 
introduce $\Sym^{J}$ with
some set of integers $J$ that indicates the operand is 
a symmetric function
of $x_{j}$, $j\in J$.
>From the identity \myref{eqn:Identity_di_and_dj}, we have
\begin{eqnarray}
  & & n_{k+1}\Sym^{[k+1]} \nonumber\\
  & & = n_{k}(d_{k+1}+ka)\Sym^{[k+1]}\nonumber\\
  & & = \Bigl\{kan_{k}+n_{k-1}d_{k+1}(d_{k}+(k-1)a)\nonumber\\
  & & \Cusr +an_{k-1}(d_{k+1}-d_{k})\Bigr\}\Sym^{[k+1]}
  \nonumber\\
  & & = \Bigl\{kan_{k}+(aK_{k\;k+1}-a)n_{k}\nonumber\\
  & & \Cusr +n_{k-1}d_{k+1}(d_{k}-(k-1)a)\Bigr\}\Sym^{[k+1]}
  \nonumber\\
  & & \vdots \nonumber\\
  & & = \Bigl\{\bigl((k-(k-1))a+a\sum_{j=2}^{k}K_{j\;k+1}\bigr)n_{k}
  \nonumber\\
  & & \Cusr 
  +d_{1}K_{1\;k+1}K_{1\;k+1}d_{k+1}(d_{2}+a)\cdots\nonumber\\
  & & \Cusr\Cusr(d_{k}+(k-1)a)\Bigr\}\Sym^{[k+1]}
  \nonumber\\
  & & = \Bigl\{d_{1}K_{1\;k+1}+a+a\sum_{j=2}^{k}K_{j\;k+1}\Bigr\}
  n_{k}\Sym^{[k+1]},
  \label{eqn:Null_4}
\end{eqnarray}
which means
\begin{equation}
  n_{k+1}\Sym^{[k+1]}\sim n_{k}\Sym^{[k+1]}.
  \label{eqn:Null_5}
\end{equation}
Thanks to the above relation, eq. \myref{eqn:Null_2} for $i>k$ 
follows from
that for $i=k$, i.e.,
\begin{equation}
  \bigl[n_{k+1},\bp_{k}\bigr]\Sym\sim n_{k+1}\Sym.
  \label{eqn:Null_6}
\end{equation}
We shall prove a stronger statement, namely,
\begin{proposition}
  \begin{equation}
    \bigl[n_{k+1}^{[M]},{\bp_{k}}^{[M]}\bigr]\Sym^{[N]}
    \sim n_{k+1}^{[M]}\Sym^{[N]},\;\;M\geq N.
    \label{eqn:Null_operator_2}
  \end{equation}
  \label{prop:Null_operator_2}
\end{proposition}
Here, the superscript $[M]$ over Dunkl operators indicates 
that they are
made from the Dunkl operators \myref{eqn:Dunkl}
that depend not only on the variables
$x_{1},x_{2},\cdots,x_{N}$ but also on $x_{N+1},\cdots,x_{M}$.
Note that $n_{k+1}^{[M]}$ and ${\bp_{k}}^{[M]}$ are 
symmetric under $S_{N}$
but not under $S_{M}$. We just changed the number of 
variables of Dunkl
operators \myref{eqn:Dunkl} but do not change 
the definition of operators
made from them, such as conserved operators
\myref{eqn:Calogero_conserved_operators_Dunkl}
and generators \myref{eqn:Parts_of_generator} and 
\myref{eqn:Generator}.
Namely, the indices and subsets in the summand are 
included in the set $[N]$.

All the Dunkl operators in
the remainder of this \S\ref{subsec:null_operators}
will always depend on
$x_{1},\cdots,x_{M}$. We shall omit the superscript $[M]$ in 
the following.

To prove Proposition \ref{prop:Null_operator_2}, 
we need several lemmas.
We define the restricted generator by
\begin{equation}
  \bp_{k,J} = 
  \sum_{\stackrel{\scriptstyle J^{\prime}\subseteq J}
  {|J^{\prime}|=k}}\dalpha_{J^{\prime}}d_{1,J^{\prime}},
  \label{eqn:Restricted_generator}
\end{equation}
where $J$ is a set of integers. 
>From the definition of generators
\myref{eqn:Generator}, we can easily verify
\begin{lemma}
  \begin{equation}
    \bp_{k}=\sum_{l=0}^{k}
    \sum_{\stackrel{\scriptstyle J\subset\{k+2,\cdots,N\}}{|J|=l}}
    \dalpha_{J}\bp_{k-l,[k+1]}d_{k-l+1,J},
    \label{eqn:Decomposition_of_generator}
  \end{equation}
  with $\bp_{0,J}=1$ and when $|J|=0$, $\dalpha_{J}=1$ 
  and $d_{k+1,J}=1$.
  \label{lemma:Decomposition_of_generator}
\end{lemma}
We shall also use the following formulae:
\begin{lemma}
  \begin{subequations}
    \begin{eqnarray}
      & & \bigl[d_{i},\dalpha_{J}\bigr] = 
      -a\dalpha_{i}\sum_{j\in J}
      \dalpha_{J\setminus \{j\}}K_{ij},\;\;i\notin J,
      \label{eqn:Null_formula_1a}\\
      & & \bigl[d_{i},\dalpha_{J}\bigr] = \dalpha_{J}\bigl(1+a
      \sum_{j\in [M]\setminus J}K_{ij}\bigr),\;\; i\in J,
      \label{eqn:Null_formula_1b}\\
      & & \bigl[n_{k+1},\dalpha_{l}\bigr]\Sym^{[k+1]}\nonumber\\
      & & \Usr = -a\bigl(\dalpha_{1}K_{12}K_{23}\cdots K_{k\;k+1}
      K_{k+1\;l}
      \nonumber\\
      & & \Usr\Usr +\cdots +\dalpha_{k+1}K_{k+1\;l}\bigr)n_{k}
      \Sym^{[k+1]}
      \nonumber\\
      & & \Usr \sim n_{k}\Sym^{[k+1]},\; l\notin [k+1].
      \label{eqn:Null_formula_1c}
    \end{eqnarray}
    \label{eqn:Null_formula_1}
  \end{subequations}
  \label{lemma:Null_formula_1}
\end{lemma}
The first two formulae can be checked from the definition 
of $d_{i}$
and $\dalpha_{J}$ and commutation relations
\myref{eqn:Needed_algebra_harmonic_Dunkl}. The last formula
\myref{eqn:Null_formula_1c} is proved by induction on $k$. 

In the following, we often need to identify the terms that 
do not have
$\dalpha_{1}$ appearing as an explicit factor on the left 
of the terms.
When ${\cal O}$ represents a Dunkl operator of our interest, 
such terms are
denoted by ${\cal O}\Bigr|_{\dalpha_{1}\sim 0}$.
\begin{lemma}
  For $M\geq n \geq k+1$, we have
  \begin{equation}
    \Bigl(\bigl[n_{k+1},\dalpha_{2}\cdots\dalpha_{n}\bigr]
    +\dalpha_{2}\cdots
    \dalpha_{n}\bigl[d_{1},d_{\{2,\cdots,k+1\}}\bigr]\Bigr)
    \Bigr|_{\dalpha_{1}\sim 0}\sim d_{1}.
    \label{eqn:Null_formula_2}
  \end{equation}
  \label{lemma:Null_formula_2}
\end{lemma}
This formula is also proved by induction on $k$.

As a step toward proving Proposition \ref{prop:Null_operator_2},
we prove the case $N=k+1$.
\begin{proposition}
  \begin{equation}
    \bigl[n_{k+1},\bp_{k,[k+1]}\bigr]\Sym^{[k+1]}\sim
    n_{k+1}\Sym^{[k+1]}.
    \label{eqn:Null_operator_3}
  \end{equation}
  \label{prop:Null_operator_3}
\end{proposition}
Both sides of the above equation are symmetric under 
permutations of indices
$1,\cdots,k+1$.
>From the first two equations of Lemma \ref{lemma:Null_formula_1}, 
we have
\begin{equation}
  \bigl[n_{k+1},\dalpha_{1}
  \cdots^{\stackrel{\scriptstyle i}{\vee}}\!\cdots
  \dalpha_{k+1}\bigr]
  = \sum_{j=1}^{N}\dalpha_{1}
  \cdots^{\stackrel{\scriptstyle j}{\vee}}\!\cdots
  \dalpha_{k+1}
  {\cal O}_{ij},
  \label{eqn:NO3_1}
\end{equation}
where ${\cal O}_{ij}$ is some unspecified operator that can 
be written by
$d_{l}$ and $K_{lm}$ with $1\leq l,m \leq k+1$ and
$\dalpha_{1}\cdots^{\stackrel{\scriptstyle i}{\vee}}\!\cdots
\dalpha_{k+1}
=\prod_{\stackrel{\scriptstyle j=1}{j\neq i}}^{k+1}\dalpha_{j}$.
Commutators among the operators made of $d_{l}$ operators are also
operators made of $d_{l}$ and $K_{lm}$ with $1\leq l,m \leq k+1$.
Thus we can say that the generator $\bp_{k,[k+1]}$ and the
l.h.s. of eq. \myref{eqn:Null_operator_3} have factors
$\dalpha_{1}\cdots^{\stackrel{\scriptstyle i}{\vee}}\!\cdots
\dalpha_{k+1}$
on the left. Therefore, in order to prove 
Proposition \ref{prop:Null_operator_3},
it is sufficient for us to prove the following expression:
\begin{equation}
  \bigl[n_{k+1},\bp_{k,[k+1]}\bigr]\Sym^{[k+1]}
  \Bigr|_{\dalpha_{1}\sim 0}
  \sim n_{k+1}\Sym^{[k+1]}\Bigr|_{\dalpha_{1}\sim 0}.
  \label{eqn:NO3_2}
\end{equation}
This is proved by induction on $k$.

We need a few more formulae to prove 
Proposition \ref{prop:Null_operator_2}.
\begin{lemma}
  For all sets of positive integers $J =\{j_{1},j_{2},\cdots,j_{l}\}$
  such that
  $J\cap\{1,2,\cdots,N\}=\emptyset$, we have
  \begin{subequations}
    \begin{eqnarray}
      & & \bigl[n_{k+1},\dalpha_{J}\bigr]\Sym^{[k+1]}
      \sim n_{k+1-l}\Sym^{[k+1]},
      \label{eqn:Null_formula_3a}\\
      & & \bigl(n_{k+1-l}d_{k+1-l,J}\Sym^{[k+1]}\bigr)
      \Sym^{[N]} \sim n_{k+1}\Sym^{[N]},
      \label{eqn:Null_formula_3b}\\
      & & \bigl[n_{k+1},d_{k+1-l,J}\bigr]\Sym^{[N]}\sim n_{k+1}\Sym^{[N]}.
      \label{eqn:Null_formula_3c}
    \end{eqnarray}
    \label{eqn:Null_formula_3}
  \end{subequations}
  \label{lemma:Null_formula_3}
\end{lemma}
These formulae can be straightforwardly verified by using 
the definitions
of operators, the third equation of Lemma \ref{lemma:Null_formula_1}
and the fundamental commutation relations among Dunkl operators
\myref{eqn:Needed_algebra_harmonic_Dunkl}.

Now we are ready to prove Proposition \ref{prop:Null_operator_2}.
We shall prove it by induction on $k$. First, we shall check the case
$k=1$. From the definition of generator \myref{eqn:Generator} and
\myref{eqn:Restricted_generator}, we have
\begin{eqnarray}
  & & \Mae\bigl[n_{2},\bp_{1}\bigr]\Sym^{[N]}\nonumber\\
  & & \!\!\!\!=\bigl[n_{2},\bp_{1,[2]}\bigr]\Sym^{[N]}
  +\sum_{i=3}^{N}\bigl[n_{2},\dalpha_{i}(d_{i}+a)\bigr]\Sym^{[N]}.
  \label{eqn:NO2_1}
\end{eqnarray}
According to Proposition \ref{prop:Null_operator_3}, 
the first term of the
r.h.s. of the above equation is confirmed to be
\begin{equation}
  \bigl[n_{2},\bp_{1,[2]}\bigr]\Sym^{[N]}
  \sim n_{2}\Sym^{[N]}.
  \label{eqn:NO2_2}
\end{equation}
The second term is calculated as
\begin{eqnarray}
  & & \Mae\bigl[n_{2},\dalpha_{i}(d_{i}+a)\bigr]\Sym^{[N]}
  \nonumber\\
  & & \!\!\!\!=\bigl[n_{2},\dalpha_{i}\bigr](d_{i}+a)\Sym^{[N]}
  +\dalpha_{i}\bigl[n_{2},(d_{i}+a)\bigr]\Sym^{[N]}.
  \label{eqn:NO2_3}
\end{eqnarray}
>From the third formula of Lemma \ref{lemma:Null_formula_3},
the second term of the r.h.s. of eq. \myref{eqn:NO2_3} satisfies
\begin{equation}
  \dalpha_{i}\bigl[n_{2},(d_{i}+a)\bigr]\Sym^{[N]}\sim n_{2}\Sym^{[N]}.
  \label{eqn:NO2_4}
\end{equation}
Using eq. \myref{eqn:Null_formula_1c}, we can verify that 
the first term of the
r.h.s. of eq. \myref{eqn:NO2_3} reduces to
\begin{eqnarray}
  \bigl[n_{2},\dalpha_{i}\bigr](d_{i}+a)\Sym^{[N]}
  & \sim & n_{1}(d_{i}+a)\Sym^{[N]}\nonumber\\
  & \sim & K_{2i}d_{1}(d_{2}+a)\Sym^{[N]}\nonumber\\
  & \sim & n_{2}\Sym^{[N]}.
  \label{eqn:NO2_5}
\end{eqnarray}
Summarizing the results, we have
\begin{equation}
  \bigl[n_{2},\bp_{1}\bigr]\Sym^{[N]}\sim n_{2}\Sym^{[N]}.
  \label{eqn:NO2_6}
\end{equation}
Thus we have confirmed that the proposition holds for $k=1$.

By inductive assumption, the proposition holds up to $k-1$.
What we like to show is that the commutator between $n_{k+1}$ 
and each term
of the decomposition of 
the generator \myref{eqn:Decomposition_of_generator}
is similar to $n_{k}$, i.e.,
\begin{equation}
  \bigl[n_{k+1},\dalpha_{J}\bp_{k-l,[k+1]}d_{k-l+1,J}\bigr]
  \Sym^{[N]}\sim n_{k+1}\Sym^{[N]}.
  \label{eqn:NO2_7}
\end{equation}
Using the Leibniz rule, we decompose the l.h.s. into two parts:
\begin{eqnarray}
  & & \Mae\bigl[n_{k+1},\dalpha_{J}\bp_{k-l,[k+1]}d_{k-l+1,J}\bigr]
  \Sym^{[N]}\nonumber\\
  & & = \bigl[n_{k+1},\dalpha_{J}\bp_{k-l,[k+1]}\bigr]d_{k-l+1,J}
  \Sym^{[N]}\nonumber\\
  & & \Cusr +\dalpha_{J}\bp_{k-l,[k+1]}\bigl[n_{k+1},d_{k-l+1,J}\bigr]
  \Sym^{[N]}.
  \label{eqn:NO2_8}
\end{eqnarray}
Then from the third formula of Lemma \ref{lemma:Null_formula_3},
we notice that the second term is similar to $n_{k+1}$:
\begin{equation}
  \dalpha_{J}\bp_{k-l,[k+1]}\bigl[n_{k+1},d_{k-l+1,J}\bigr]
  \Sym^{[N]}\sim n_{k+1}\Sym^{[N]}.
  \label{eqn:NO2_9}
\end{equation}
Our remaining task is to check the first term. When $l=|J|=0$,
the first term
is similar to $n_{k+1}$ because of 
Proposition \ref{prop:Null_operator_3}.
When $l\neq 0$, we have to do some calculation:
\begin{eqnarray}
  & & \Mae \bigl[n_{k+1},\dalpha_{J}\bp_{k-l,[k+1]}\bigr]
  d_{k-l+1,J}\Sym^{[N]}\nonumber\\
   & & = \bigl[n_{k+1},\dalpha_{J}\bigr]\bp_{k-l,[k+1]}
  d_{k-l+1,J}\Sym^{[N]}\nonumber\\
  & & \Cusr + \dalpha_{J}\bigl[n_{k+1},\bp_{k-l,[k+1]}\bigr]
  d_{k-l+1,J}\Sym^{[N]}.
  \label{eqn:NO2_10}
\end{eqnarray}
>From the first formula of Lemma \ref{lemma:Null_formula_3}, 
the first term of the r.h.s. of eq. \myref{eqn:NO2_10} 
is calculated as
\begin{eqnarray}
  & & \Mae \bigl[n_{k+1},\dalpha_{J}\bigr]\bp_{k-l,[k+1]}
  d_{k-l+1,J}\Sym^{[N]}\nonumber\\
  % & & = \bigl[n_{k+1},\dalpha_{J}\bigr]\Sym^{[k+1]}\bp_{k-l,[k+1]}
  % d_{k-l+1,J}\Sym^{[N]}\nonumber\\
  & & \sim n_{k+1-l}\bp_{k-l,[k+1]}d_{k-l+1,J}\Sym^{[N]}
  \nonumber\\
  & & = \bp_{k-l,[k+1]}n_{k+1-l}d_{k-l+1,J}\Sym^{[N]}
  \nonumber\\
  & & \Cusr + \bigl[n_{k+1-l},\bp_{k-l,[k+1]}\bigr]d_{k-l+1,J}
  \Sym^{[N]}.
  \label{eqn:NO2_11}
\end{eqnarray}
Note that the operators $\bp_{k-l,[k+1]}$ and 
$d_{k-l+1,J}$ are invariant under the permutations of indices
$1,\cdots,k+1$. Using the second formula of 
Lemma \ref{lemma:Null_formula_3},
we can verify that the first term of the r.h.s. of 
eq. \myref{eqn:NO2_11}
reduces to
\begin{equation}
  \bp_{k-l,[k+1]}n_{k+1-l}d_{k-l+1,J}\Sym^{[N]}\sim n_{k+1}\Sym^{[N]}.
  \label{eqn:NO2_12}
\end{equation}
>From the induction hypothesis, we have
\begin{equation}
  \bigl[n_{k-l+1},\bp_{k-l,[k+1]}\bigr]\Sym^{[k+1]}\sim
  n_{k-l+1}\Sym^{[k+1]}.
  \label{eqn:NO2_13}
\end{equation}
Then the second term of  the r.h.s. of eq. \myref{eqn:NO2_11} 
is calculated as
\begin{eqnarray}
  & & \Mae \bigl[n_{k+1-l},\bp_{k-l,[k+1]}\bigr]\Sym^{[k+1]}
  d_{k-l+1,J}\Sym^{[N]}\nonumber\\
  & & \sim n_{k-l+1}d_{k-l+1,J}\Sym^{[N]}\nonumber\\
  & & \sim n_{k+1}\Sym^{[N]}.
  \label{eqn:NO2_14}
\end{eqnarray}
On the other hand,
the second term of the r.h.s. of eq. \myref{eqn:NO2_10} 
is separated into
two parts as
\begin{eqnarray}
  & & \Mae \dalpha_{J}\bigl[n_{k+1},\bp_{k-l,[k+1]}\bigr]
  d_{k-l+1,J}\Sym^{[N]}\nonumber\\
  & & = \dalpha_{J}n_{k+1}\bp_{k-l,[k+1]}d_{k-l+1,J}\Sym^{[N]}
  \nonumber\\
  & & \Cusr -\dalpha_{J}\bp_{k-l,[k+1]}n_{k+1}d_{k-l+1,J}\Sym^{[N]}.
  \label{eqn:NO2_15}
\end{eqnarray}
In an analogous way to the verification of
the first and the second terms in the r.h.s. of 
eq. \myref{eqn:NO2_11},
both the second and the first terms in the r.h.s. of 
eq. \myref{eqn:NO2_15}
are respectively similar to $n_{k+1}$. Then we obtain
\begin{equation}
  \dalpha_{J}\bigl[n_{k+1},\bp_{k-l,[k+1]}\bigr]
  d_{k-l+1,J}\Sym^{[N]}\sim n_{k+1}\Sym^{[N]}.
  \label{eqn:NO2_16}
\end{equation}
Equations \myref{eqn:NO2_12}, \myref{eqn:NO2_14} 
and \myref{eqn:NO2_16}
yield
\begin{equation}
  \bigl[n_{k+1},\dalpha_{J}\bp_{k-l,[k+1]}\bigr]
  d_{k-l+1,J}\Sym^{[N]}\sim n_{k+1}\Sym^{[N]}.
  \label{eqn:NO2_17}
\end{equation}
Thus we have proved Proposition \ref{prop:Null_operator_2}.

\subsection{Generators}\label{subsec:generators}
Proposition \ref{prop:Eigenvalue_precursor} is given in 
a form whose operators
depend on $N$ variables $x_{1},\cdots,x_{N}$.
It is convenient for us
to explicitly indicate the number of variables, 
for we shall change the number
of variables during the induction procedure.
\begin{proposition}
  \begin{eqnarray}
    & & \Mae \bigl[I_{2}^{[N]}(N),\bpr{N}_{k,[N]}\bigr]\Sym^{[N]}
    \nonumber\\
    & & = \Bigl\{\bpr{N}_{k,[N]}\bigl(2I_{1}^{[N]}(N)+k+ak(N-k)\bigr)
    \nonumber\\
    & & \Usr +\sum_{\stackrel{\scriptstyle J\subseteq [N]}{|J|=k+1}}
    g_{k+1,J}^{[N]}n_{k+1,J}^{[N]}\Bigr\}\Sym^{[N]},\; N\geq k,
    \label{eqn:EP}
  \end{eqnarray}
  where $g_{k+1,J}^{[N]}$ is an unspecified 
  nonsingular operator that satisfies
  $g_{N+1,J}^{[N]}=0$.
  \label{prop:EP}
\end{proposition}
Note that we have introduced the notation
\begin{equation}
  I_{n}^{[M]}(N)=\sum_{i=1}^{N}(d_{i}^{[M]})^{n}.
  \label{eqn:ND_InM}
\end{equation}
It is obvious that 
Proposition \ref{prop:Eigenvalue_precursor} is tantamount to
the above proposition.
We shall prove Proposition \ref{prop:EP} by induction on
$k$ and $N$. Precisely speaking, we use the induction on $l$, 
which relates $k$
and $N$ by $k=l+1$ and $N=l+M$ with arbitrary integer $M$.
The plan requires several lemmas.
\begin{lemma}
  \begin{subequations}
    \begin{eqnarray}
      & & \Mae \bigl[(d_{j}^{[M]})^{2},\nkdr{1}{i}{N}{M}\bigr]
      \nonumber\\
      & & = \nkdr{1}{i}{N}{M}\nonumber\\
      & & \Usr \bigl\{(1+aK_{ij}+a\sum_{k=N+1}^{M}K_{jk})^{2}
      \nonumber\\
      & & \Usr +(1+aK_{ij}+a\sum_{k=N+1}^{N}K_{jk})d_{j}^{[M]}
      \nonumber\\
      & & \Usr
      +d_{j}^{[M]}(1+aK_{ij}+a\sum_{k=N+1}^{N}K_{jk})\bigr\},
      \nonumber\\
      & & \Usr\Usr\Usr 1\leq i,j\leq N,\; i\neq j,
      \label{eqn:EP_formula_1a}\\
      & & \Mae \bigl[(d_{j}^{[M]})^{2},
      \nkdr{1}{i}{N}{M}\bigr]\dalzero{1}\nonumber\\
      & & = -a\nkdr{2}{i}{N}{M}{\dalpha_{j}}^{[M]}\nonumber\\
      & & \Cusr\bigl(d_{j}K_{1j}+K_{1j}d_{j}+K_{1j}\nonumber\\
      & & \Cusr +aK_{1j}(K_{1i}+\sum_{k=N+1}^{M}K_{1k})\bigr),
      \nonumber\\
      & & \Usr\Usr\Usr j=i\;\mbox{\rm or }N+1\leq j\leq M.
      \label{eqn:EP_formula_1b}
    \end{eqnarray}
    \label{eqn:EP_formula_1}
  \end{subequations}
  \label{lemma:EP_formula_1}
\end{lemma}
Both formulae are readily verified from the first two formulae of
Lemma \ref{lemma:Null_formula_3}.
Lemma \ref{lemma:EP_formula_1} leads to the following formula.
\begin{corollary}
  For $1\leq i<N\leq M$, we have
  \begin{eqnarray}
    & & \bigl[I_{2}^{[M]}(N-1),\nkdr{1}{i}{N-1}{M}
    {\dalpha_{k}}^{[M]}\bigr]\dalzero{1}\nonumber\\
    & & \;\; = {\dalpha_{k}}^{[M]}\bigl[I_{2}^{[M-1]}(N-1),
    \nkdr{1}{i}{N-1}{M-1}\bigr]
    \nonumber\\
    & & \Usr \dalzero{1}\Bigr|_{{\cal O}^{[M-1]}\sim{\cal O}^{[M]}},\; 
    N\leq k\leq M.
    \label{eqn:EP_formula_2}
  \end{eqnarray}
  \label{cor:EP_formula_2}
\end{corollary}
The restriction $\Bigr|_{{\cal O}^{[M-1]}\sim{\cal O}^{[M]}}$ 
means that we
respectively identify the Dunkl operators 
${\dalpha_{i}}^{[M]},\alpha_{i}^{[M]}$ and $d_{i}^{[M]}$ with
${\dalpha_{i}}^{[M-1]},\alpha_{i}^{[M-1]}$ and $d_{i}^{[M-1]}$,
where $i\leq M-1$.

The following formulae are valid for arbitrary number of variables.
\begin{lemma}
  For $M \geq N$, we have
  \begin{subequations}
    \begin{eqnarray}
      & & \Mae \;\;\bigl[I_{n}^{[M]}(N-1),d_{N}^{[M]}\bigr]
      \Sym^{[N]}
      \nonumber\\
      & & = a\bigl((N-1)(d_{N}^{[M]})^{n}-I_{n}^{[M]}(N-1)\bigr)
      \Sym^{[N]},
      \label{eqn:EP_formula_3a}\\
      & & \Mae \;\;
      \bigl[(d_{N}^{[M]})^{n},d_{1,[N]}^{[M]}\bigr]\Sym^{[N]}
      \nonumber\\
      & & =ad_{1,[N-1]}^{[M]}
      \bigl(I_{n}^{[M]}(N-1)-(N-1)(d_{N}^{[M]})^{n}\bigr)\Sym^{[N]},
      \nonumber\\
      & & \Usr\Usr \forall N\geq 2.
      \label{eqn:EP_formula_3b}
    \end{eqnarray}
    \label{eqn:EP_formula_3}
  \end{subequations}
  \label{lemma:EP_formula_3}
\end{lemma}
The first identity can be checked from eq.
\myref{eqn:Commutator_harmonic_Dunkl_2}. The second one is 
proved by
induction on $N$ from again 
eq. \myref{eqn:Commutator_harmonic_Dunkl_2}.

Now we shall start the proof of Proposition \ref{prop:EP}.
We have to separately prove the case for $k=N$ because 
of the difference of
the definition of the generator \myref{eqn:Generator}:
\begin{proposition}
  \begin{equation}
    \bigl[I_{2}^{[N]}(N),\bpr{N,[N]}_{N}\bigr]
    =\bpr{N,[N]}_{N}\bigl(2I_{1}^{[N]}(N)+N)\bigr).
    \label{eqn:EP_special}
  \end{equation}
\end{proposition}
This formula can be straightforwardly verified from 
the definition
of $\bpr{N}_{N,[N]}$ \myref{eqn:Generator_edge} and 
the second formula of
Lemma \ref{lemma:Null_formula_1}.

As a ground for inductive assumption, 
we need a proposition:
\begin{proposition}
  \begin{eqnarray}
    & & \Mae\bigl[I_{2}^{[M]}(M),{\bpr{M}_{1,[M]}}\bigr]\Sym^{[M]}
    \nonumber\\
    & & = \Bigl(\bigl({\bpr{M}_{1,[M]}}I_{1}^{[M]}(M)+1+a(M-1)\bigr)
    \nonumber\\
    & & \Usr +\sum_{\stackrel{\scriptstyle J\subseteq [M]}{|J|=2}}
    g_{2,J}^{[M]}n_{2,J}^{[M]}\Bigr)\Sym^{[M]},\;\forall M\geq 1.
    \label{eqn:EP_ground}
  \end{eqnarray}
  \label{prop:EP_ground}
\end{proposition}
This is nothing but 
Proposition \ref{prop:EP} for $l=0$, i.e., $k=1$ and
$N=M\geq 1$.
The proof is as follows. 
Using eq. \myref{eqn:Commutator_harmonic_Dunkl_2},
we can rewrite the second conserved operator $I_{2}^{[M]}(M)$ as
\begin{eqnarray}
  & & \!\!I_{2}^{[M]}(M)\Sym^{[M]}\nonumber\\
  & & = \Bigl((d_{1}^{[M]}+\cdots+d_{M}^{[M]})^{2}\nonumber\\
  & & \Usr -\sum_{1\leq i<j\leq M}
  (d_{i}^{[M]}d_{j}^{[M]}+d_{j}^{[M]}d_{i}^{[M]})\Bigr)\Sym^{[M]}
  \nonumber\\
  & & = \Bigl(\bigl(d_{1}^{[M]}+\cdots+d_{M}^{[M]}\bigr)
  \bigl(d_{1}^{[M]}+\cdots+d_{M}^{[M]}+a(M-1)\bigr)\nonumber\\
  & & \Usr -2\sum_{1\leq i<j\leq M}d_{i}^{[M]}(d_{j}^{[M]}+a)\Bigr)
  \Sym^{[M]}
  \nonumber\\
  & & = \Bigl(I_{1}^{[M]}(M)\bigl(I_{1}^{[M]}(M)+a(M-1)\bigr)
  \nonumber\\
  & & \Usr -2\sum_{\stackrel{\scriptstyle J\subseteq [M]}{|J|=2}}
  n_{2,J}^{[M]}\Bigr)\Sym^{[M]}.
  \label{eqn:EPG_1}
\end{eqnarray}
Then from Proposition \ref{prop:Homogeneous} and Proposition
\ref{prop:Null_operator_2} for $k=1$, i.e., eq. \myref{eqn:NO2_6},
we obtain the results.

We assume that Proposition \ref{prop:EP} is valid up to $l$,
namely, $k=l+1$ and $N=l+M$. Assuming the validity of 
Proposition \ref{prop:EP} for the case of $k$ and $N$
we shall verify the case of $k+1$ and $N+1$:
\begin{eqnarray}
  & & \Mae \bigl[I_{2}^{[N+1]}(N+1),\bpr{N+1}_{k+1,[N+1]}\bigr]
  \Sym^{[N+1]}
  \nonumber\\
  & & = \Bigl\{\bpr{N+1}_{k+1,[N+1]}\bigl(2I_{1}^{[N+1]}(N+1)+(k+1)
  \nonumber\\
  & & \Usr +a(k+1)(N-k)\bigr)
  \nonumber\\
  & & \Usr +\sum_{\stackrel{\scriptstyle J\subseteq [N+1]}{|J|=k+2}}
  g_{k+2,J}^{[N+1]}n_{k+2,J}^{[N+1]}\Bigr\}\Sym^{[N+1]}.
  \label{eqn:EPP_1}
\end{eqnarray}
Since both sides of the above expression are invariant 
under the permutations
of indices $1,2,\cdots,N+1$, it is sufficient to check the term
with the factor 
$\dalphar{N+1}_{2}\cdots\dalphar{N+1}_{k+1}\dalphar{N+1}_{N+1}$
on the left. Thus we introduce the restriction 
$\dalphar{N+1}_{1},\dalphar{N+1}_{k+2},\cdots,\dalphar{N+1}_{N}\sim 0$,
which we denote by
$\bigr|_{\dalphar{N+1}_{1,k+2,\cdots,N}\sim 0}$.
>From the definition of the conserved operator $I^{[N]}_{2}(N)$ 
\myref{eqn:ND_InM} and
Lemma \ref{lemma:Decomposition_of_generator}, we have
\begin{eqnarray}
  I^{[N+1]}_{2}(N+1) & = & I^{[N+1]}_{2}(N)+(d_{N+1}^{[N+1]})^{2},
  \label{eqn:Decomposition_I2N+1}\\
  \bpr{N+1}_{k+1,[N+1]} & = & \dalphar{N+1}_{N+1}\bpr{N+1}_{k,[N]}
  \bigl(d^{[N+1]}_{N+1}+(k+1)a\bigr)\nonumber\\
  & & +\bpr{N+1}_{k+1,[N]}.
  \label{eqn:Decomposition_BPRN+1}
\end{eqnarray}
Then the l.h.s. of eq. \myref{eqn:EPP_1} is decomposed as
\begin{eqnarray}
  & & \!\!\bigl[I_{2}^{[N+1]}(N+1),
  \bpr{N+1}_{k+1,[N+1]}\bigr]\Rest\Sym^{[N+1]}
  \nonumber\\
  & & = \Bigl\{\bigl[I_{2}^{[N+1]}(N),
  \dalphar{N+1}_{N+1}\bpr{N+1}_{k,[N]}
  \bigl(d^{[N+1]}_{N+1}+(k+1)a\bigr)\bigr]\nonumber\\
  & & \Usr + \bigl[I_{2}^{[N+1]}(N),\bpr{N+1}_{k+1,[N]}\bigr]
  \nonumber\\
  & & \Usr + \bigl[(d_{N+1}^{[N+1]})^{2},\dalphar{N+1}_{N+1}
  \bpr{N+1}_{k,[N]}\bigl(d^{[N+1]}_{N+1}+(k+1)a\bigr)\bigr]
  \nonumber\\
  & & \Usr + \bigl[(d_{N+1}^{[N+1]})^{2},\bpr{N+1}_{k+1,[N]}\bigr]
  \Bigr\}\nonumber\\
  & & \Cusr\Usr\Rest\Sym^{[N+1]}.
  \label{eqn:EPP_2}
\end{eqnarray}
Because of Lemma \ref{lemma:EP_formula_1}, we can easily confirm
\begin{subequations}
  \begin{eqnarray}
    & & 
    \Cmae\bigl[I_{2}^{[N+1]}(N),\bpr{N+1}_{k+1,[N]}\bigr]
    \Rest\Sym^{[N+1]}=0,
    \label{eqn:EPP_3a}\\
    & & 
    \Cmae\bigl[(d_{N+1}^{[N+1]})^{2},\dalphar{N+1}_{N+1}
    \bpr{N+1}_{k,[N]}\bigr]\Rest\Sym^{[N+1]}\nonumber\\
    & & = \bigl[(d_{N+1}^{[N+1]})^{2},\dalphar{N+1}_{N+1}
    \dalphar{N+1}_{2}\cdots
    \dalphar{N+1}_{k+1}d^{[N+1]}_{1,\{2,\cdots,k+1\}}
    \bigr]\nonumber\\
    & & \Usr \Rest\Sym^{[N+1]},
    \label{eqn:EPP_3b}\\
    & & \Cmae\bigl[(d_{N+1}^{[N+1]})^{2},\bpr{N+1}_{k+1,[N]}\bigr]
    \Rest\Sym^{[N+1]}\nonumber\\
    & & = \sum_{i\in\{1,k+2,\cdots,N\}}\bigl[(d_{N+1}^{[N+1]})^{2},
    \dalphar{N+1}_{i}\dalpha_{2}\cdots\dalphar{N+1}_{k+1}\bigr]
    \nonumber\\
    & & \Usr d^{[N+1]}_{1,\{2,\cdots,k+1\}}
    \bigl(d^{[N+1]}_{i}+(k+1)a\bigr)
    \nonumber\\
    & & \Usr \Rest\Sym^{[N+1]}.
    \label{eqn:EPP_3c}
  \end{eqnarray}
  \label{eqn:EPP_3}
\end{subequations}
Thus the r.h.s. of eq. \myref{eqn:EPP_2} becomes
\begin{eqnarray}
  & & \!\!\bigl[I_{2}^{[N+1]}(N+1),\bpr{N+1}_{k+1,[N+1]}\bigr]
  \nonumber\\
  & & = \Bigl\{\bigl[I_{2}^{[N+1]}(N),\dalphar{N+1}_{N+1}
  \bpr{N+1}_{k,[N]}\bigr]\bigl(d^{[N+1]}_{N+1}+(k+1)a\bigr)
  \nonumber\\
  & & \Cusr +\dalphar{N+1}_{N+1}\dalphar{N+1}_{2}\cdots
  \dalphar{N+1}_{k+1}
  d^{[N+1]}_{1,\{2,\cdots,k+1\}}\nonumber\\
  & & \Usr \bigl[I_{2}^{[N+1]}(N),d^{[N+1]}_{N+1}\bigr]
  \nonumber\\
  & & \Cusr + \bigl[(d_{N+1}^{[N+1]})^{2},\dalphar{N+1}_{N+1}
  \dalphar{N+1}_{2}\cdots\dalphar{N+1}_{k+1}
  \nonumber\\
  & & \Cusr\Usr d^{[N+1]}_{1,\{2,\cdots,k+1,N+1\}}\bigr]
  \nonumber\\
  & & \Cusr + \sum_{i\in\{1,k+2,\cdots,N\}}
  \bigl[(d_{N+1}^{[N+1]})^{2},
  \dalphar{N+1}_{i}\dalpha_{2}\cdots\dalphar{N+1}_{k+1}\bigr]
  \nonumber\\
  & & \Usr d^{[N+1]}_{1,\{2,\cdots,k+1\}}
  \bigl(d^{[N+1]}_{i}+(k+1)a\bigr)
  \Bigr\}\nonumber\\
  & & \Usr \Rest\Sym^{[N+1]}.
  \label{eqn:EPP_4}
\end{eqnarray}
We shall calculate the above equation term by term.
Using Corollary \ref{cor:EP_formula_2},
we can rewrite the commutator of 
the first term of the r.h.s. of eq. \myref{eqn:EPP_4} as
\begin{eqnarray}
  & & \Mae \bigl[I_{2}^{[N+1]}(N),\dalphar{N+1}_{N+1}
  \bpr{N+1}_{k,[N]}\bigr]\nonumber\\
  & & \bigl(d^{[N+1]}_{N+1}+(k+1)a\bigr)\Rest\Sym^{[N+1]}
  \nonumber\\
  & & = \dalphar{N+1}_{N+1}
  \bigl[I_{2}^{[N]}(N),\bpr{N}_{k,[N]}\bigr]
  \Sym^{[N]}\Bigr|_{{\cal O}^{[N]}\sim{\cal O}^{[N+1]}}
  \nonumber\\
  & & \Usr \bigl(d^{[N+1]}_{N+1}+(k+1)a\bigr)\Rest\Sym^{[N+1]}.
  \label{eqn:EPP_5}
\end{eqnarray}
The above commutator allows the use of inductive assumption. 
Thus we have
\begin{eqnarray}
  & & \Cmae \bigl[I_{2}^{[N+1]}(N),\dalphar{N+1}_{N+1}
  \bpr{N+1}_{k,[N]}\bigr]\nonumber\\
  & & \bigl(d^{[N+1]}_{N+1}+(k+1)a\bigr)\Rest\Sym^{[N+1]}
  \nonumber\\
  & & = \Bigl\{\dalphar{N+1}_{2}\cdots\dalphar{N+1}_{k+1}
  \dalphar{N+1}_{N+1}
  d^{[N+1]}_{1,\{2,\cdots,k+1,N+1\}}\nonumber\\
  & & \Usr\bigl(2I_{1}^{[N+1]}(N)+k+ak(N-k)\bigr)\nonumber\\
  & & \Cusr + 2a\dalphar{N+1}_{2}\cdots
  \dalphar{N+1}_{k+1}\dalphar{N+1}_{N+1}
  d^{[N+1]}_{1,\{2,\cdots,k+1\}}\nonumber\\
  & & \Usr \bigl(Nd^{[N+1]}_{N+1}-I_{1}^{[N+1]}(N)\bigr)
  \nonumber\\
  & & \Cusr + \dalphar{N+1}_{N+1}\sum_{\stackrel
  {\scriptstyle J\subseteq [k]\cup\{N+1\}}{N+1\in J,|J|=k+2}}
  g_{N-1,J\setminus\{N+1\}}^{[N+1]}n^{[N+1]}_{k+2,J}\Bigr\}
  \nonumber\\
  & & \Usr \Rest\Sym^{[N+1]}.
  \label{eqn:EPP_6}
\end{eqnarray}
The second term is straightforwardly calculated from eq.
\myref{eqn:EP_formula_3a}:
\begin{eqnarray}
  & & \Mae\;\; \dalphar{N+1}_{N+1}\dalphar{N+1}_{2}\cdots
  \dalphar{N+1}_{k+1}
  d^{[N+1]}_{1,\{2,\cdots,k+1\}}\nonumber\\
  & & \bigl[I_{2}^{[N+1]}(N),d^{[N+1]}_{N+1}\bigr]\Rest\Sym^{[N+1]}
  \nonumber\\
  & & = a\dalphar{N+1}_{N+1}\dalphar{N+1}_{2}\cdots
  \dalphar{N+1}_{k+1}
  d^{[N+1]}_{1,\{2,\cdots,k+1\}}\nonumber\\
  & & \Cusr \bigl(N(d^{[N+1]}_{N+1})^{2}
  -I^{[N+1]}_{2}(N)\bigr)\Rest\Sym^{[N+1]}.\nonumber\\
  \label{eqn:EPP_7}
\end{eqnarray}
By using eqs. \myref{eqn:EP_formula_1a} and 
\myref{eqn:EP_formula_3b},
the third term is cast into
\begin{eqnarray}
  & & \! \bigl[(d^{[N+1]}_{N+1})^{2},\dalphar{N+1}_{2}\cdots
  \dalphar{N+1}_{k+1}\dalphar{N+1}_{N+1}
  d^{[N+1]}_{1,\{2,\cdots,k+1,N+1\}}\bigr]
  \nonumber\\
  & & \Cusr\Rest\Sym^{[N+1]}\nonumber\\
  & & = \Bigl\{\bigl[(d^{[N+1]}_{N+1})^{2},\dalphar{N+1}_{2}\cdots
  \dalphar{N+1}_{k+1}\dalphar{N+1}_{N+1}\bigr]\nonumber\\
  & & \Usr d^{[N+1]}_{1,\{2,\cdots,k+1,N+1\}}
  \nonumber\\
  & & \Cusr + \dalphar{N+1}_{2}\cdots\dalphar{N+1}_{k+1}
  \dalphar{N+1}_{N+1}
  \nonumber\\
  & & \Usr \bigl[(d^{[N+1]}_{N+1})^{2},
  d^{[N+1]}_{1,\{2,\cdots,k+1,N+1\}}
  \bigr]\Bigr\}\Rest\Sym^{[N+1]}\nonumber\\
  & & = \dalphar{N+1}_{2}\cdots\dalphar{N+1}_{k+1}
  \dalphar{N+1}_{N+1}
  \nonumber\\
  & & \Cusr \Bigl\{\Bigl(2d^{[N+1]}_{N+1}+1+a^{2}(N-k)
  \nonumber\\
  & & \Usr +a\sum_{i\in\{1,k+2,\cdots,N\}}
  \bigl(2K_{i N+1}+d^{[N+1]}_{N+1}K_{i N+1}
  \nonumber\\
  & & \Usr\Cusr +K_{i N+1}d^{[N+1]}_{N+1}\bigr)
  \nonumber\\
  & & \Usr + a^{2}
  \sum_{\stackrel{\scriptstyle i,j\in\{1,k+2,\cdots,N\}}
  {i\neq j}}K_{i N+1}K_{j N+1}\Bigr)
  d^{[N+1]}_{1,\{2,\cdots,k+1,N+1\}}
  \nonumber\\
  & & \Usr + ad^{[N+1]}_{1,\{2,\cdots,k+1\}}\bigl(\sum_{i=2}^{k+1}
  (d^{[N+1]}_{i})^{2}-k(d^{[N+1]}_{N+1})^{2}\bigr)\Bigr\}
  \nonumber\\
  & & \Usr\Usr \Rest\Sym^{[N+1]}.
  \label{eqn:EPP_8}
\end{eqnarray}
Using eq. \myref{eqn:EP_formula_1b}, 
we get the following expression 
from the fourth term:
\begin{eqnarray}
  & & \!\! \sum_{i\in\{1,k+2,\cdots,N\}}\bigl[(d^{[N+1]}_{N+1})^{2},
  \dalphar{N+1}_{i}\dalphar{N+1}_{2}\cdots\dalphar{N+1}_{k+1}\bigr]
  \nonumber\\
  & & d^{[N+1]}_{1,\{2,\cdots,k+1,N+1\}}\bigl(d^{[N+1]}_{i}+(k+1)a\bigr)
  \Rest\Sym^{[N+1]}\nonumber\\
  & & = -a \sum_{i\in\{1,k+2,\cdots,N\}}
  \dalphar{N+1}_{2}\cdots\dalphar{N+1}_{k+1} \dalphar{N+1}_{N+1}
  \nonumber\\
  & & \Usr \bigl(K_{i N+1}d^{[N+1]}_{N+1}+d^{[N+1]}_{N+1}
  K_{i N+1}+K_{i N+1}
  \nonumber\\
  & & \Usr 
  + a + 
  a\sum_{\stackrel{\scriptstyle j\in\{1,k+2,\cdots,N\}}{j\neq i}}
  K_{j N+1}K_{i N+1}\bigr)\nonumber\\
  & & 
  \Usr d^{[N+1]}_{1,\{2,\cdots,k+1\}}\bigl(d^{[N+1]}_{i}+(k+1)a\bigr)
  \Rest\Sym^{[N+1]}.\nonumber\\
  \label{eqn:EPP_9}
\end{eqnarray}
Assembling eqs. \myref{eqn:EPP_6} -- \myref{eqn:EPP_9} and doing some
calculation with the help of Lemma \ref{lemma:EP_formula_3} and the
definition of null operators \myref{eqn:Null_operator}, we obtain
\begin{eqnarray}
  & & \bigl[I_{2}^{[N+1]}(N+1),\bpr{N+1}_{k+1,[N+1]}\bigr]
  \Rest\Sym^{[N+1]}
  \nonumber\\
  & & \;\;
  = \Bigl\{\dalphar{N+1}_{2}\cdots
  \dalphar{N+1}_{k+1} \dalphar{N+1}_{N+1}
  d^{[N+1]}_{1,\{2,\cdots,k+1,N+1\}}\nonumber\\
  & & \Usr\bigl(2I_{1}^{[N+1]}(N+1)+(k+1)+a(k+1)(N-k)\bigr)
  \nonumber\\
  & & \Usr +
  \sum_{\stackrel{\scriptstyle J\subseteq[N+1]}{N+1\in J,|J|=k+2}}
  g^{\prime[N+1]}_{k+2,J}n^{[N+1]}_{k+2,J}\Bigr\}\Rest\Sym^{[N+1]}
  \nonumber\\
  & & \;\; = \Bigl\{\bpr{N+1}_{k+1}\nonumber\\
  & & \Usr\bigl(2I_{1}^{[N+1]}(N+1)+(k+1)+a(k+1)(N-k)\bigr)
  \nonumber\\
  & & \Usr +\sum_{\stackrel{\scriptstyle J\subseteq[N+1]}{|J|=k+2}}
  g^{[N+1]}_{k+2,J}n^{[N+1]}_{k+2,J}\Bigr\}\Rest\Sym^{[N+1]},
  \label{eqn:EPP_10}
\end{eqnarray}
which proves Proposition \ref{prop:EP}.

\subsection{Normalization and triangularity}\label{subsec:normalization}
We shall prove the last two properties of the Hi-Jack polynomials 
\myref{eqn:Hi-Jack_triangularity} and 
\myref{eqn:Hi-Jack_normalization} and Propositions
\ref{prop:Macdonald-Stanley_Hi-Jack} and \ref{prop:Jack_to_Hi-Jack}.
Here again we implicitly assume that the 
Dunkl operators depend only on $N$
variables $x_{1},\cdots,x_{N}$. The restriction symbol 
without explicit
indication of indices also means the restriction 
to symmetric functions
of indices $1,2,\cdots,N$.

First we shall prove the following proposition.
\begin{proposition}
  Operation of symmetric polynomials of Dunkl operators
  $\{\dalpha_{1},\dalpha_{2},\cdots,\dalpha_{N}\}$ 
  on any symmetric
  polynomials of $\{x_{1},x_{2},\cdots,x_{N}\}$ 
  yields symmetric polynomials
  of $\{x_{1},x_{2},\cdots,x_{N}\}$.
  \label{prop:Symmetric_Dunkl_symmetric_polynomial}
\end{proposition}
Let us denote an arbitrary polynomial of $x_{1},x_{2},\cdots,x_{N}$
by $P(x_{1},x_{2},\cdots,x_{N})$. 
Acting the Dunkl operator $\dalpha_{j}$
on $P(x_{1},x_{2},\cdots,x_{N})$, we have
\begin{eqnarray}
  & & \dalpha_{j}P(x_{1},x_{2},\cdots,x_{N})\nonumber\\
  & & = \Bigl(x_{j}-\frac{1}{2\omega}
  \frac{\partial}{\partial x_{j}}\Bigr)
  P(x_{1},x_{2},\cdots,x_{N})\nonumber\\
  & & \Cusr +\frac{a}{2\omega}
  \sum^{N}_{\stackrel{\scriptstyle i=1}{i\neq j}}
  \frac{1}{x_{i}-x_{j}}\bigl(P(x_{1},x_{2},\cdots,x_{N})
  -P(x_{i}\leftrightarrow x_{j})\bigr).\nonumber\\
  \label{eqn:PP_1}
\end{eqnarray}
It is obvious that the first term of 
the r.h.s. of eq. \myref{eqn:PP_1} is
a polynomial of $x_{1},x_{2},\cdots,x_{N}$. 
Since the difference of polynomials
$P(x_{1},x_{2},\cdots,x_{N})-P(x_{i}\leftrightarrow x_{j})$ has
a zero at $x_{i}=x_{j}$, the second term is also a
polynomial of $x_{1},x_{2},\cdots,x_{N}$.
Proposition \ref{prop:Symmetric_Dunkl_symmetric_polynomial} 
follows from this
property.

As a basis of symmetric polynomials, 
we employ the monomial symmetric
polynomials~\cite{Stanley_1,Macdonald_1} defined by
\begin{equation}
  m_{\lambda}(x_{1},x_{2},\cdots,x_{N})=\sum_{
  \stackrel{\scriptstyle \sigma:\;\mbox{\scriptsize distinct}}
  {\mbox{\scriptsize permutation}}}x_{\sigma(1)}^{\lambda_{1}}
  x_{\sigma(2)}^{\lambda_{2}}\cdots x_{\sigma(N)}^{\lambda_{N}},
  \label{eqn:Monomial}
\end{equation}
where $\lambda$ is a Young tableau \myref{eqn:Young_tableau}.
By definition, monomial symmetric polynomials 
are symmetric polynomials
with respect to any exchange of indices $1,2,\cdots,N$.
>From eqs. \myref{eqn:PP_1} and \myref{eqn:Monomial},
we have a following result
as a special case of 
Proposition \ref{prop:Symmetric_Dunkl_symmetric_polynomial}:
\begin{corollary}
  \begin{eqnarray}
    & & \Mae m_{\lambda}(\dalpha_{1},\dalpha_{2},\cdots,\dalpha_{N})
    \cdot 1\nonumber\\
    & & = m_{\lambda}(x_{1},x_{2},\cdots,x_{N})\nonumber\\
    & & \Cusr + 
    \sum_{\stackrel{\scriptstyle \mu}{|\mu|<|\lambda|}}
    u_{\lambda\mu}(a,\frac{1}{2\omega})
    m_{\mu}(x_{1},x_{2},\cdots,x_{N}),
    \label{eqn:Monomial_Dunkl_Monomial_x_expansion}
  \end{eqnarray}
  where an unspecified coefficient $u_{\lambda\mu}(a,1/2\omega)$ 
  is an integer coefficient polynomial of $a$ and $1/2\omega$.
  \label{cor:Monomial_Dunkl_Monomial_x_expansion}
\end{corollary}

Next we shall consider the action of $d_{i}$ 
operator on the monomial symmetric
polynomials of $\dalpha_{1},\dalpha_{2},\cdots,\dalpha_{N}$.
We shall consider the case where the length of the 
Young tableau $\lambda$
is $l\leq N$, i.e., $\lambda=\{\lambda_{1}\geq\lambda_{2}\geq
\cdots
\geq\lambda_{l}>0\}$. From the monomial symmetric polynomial
$m_{\lambda}(\dalpha_{1},\dalpha_{2},\cdots,\dalpha_{N})$,
we single out a monomial $(\dalpha_{\sigma(1)})^{\lambda_{1}}
(\dalpha_{\sigma(2)})^{\lambda_{2}}
\cdots(\dalpha_{\sigma(l)})^{\lambda_{l}}$.
Because of eq. \myref{eqn:Action_harmonic_Dunkl}, we have
\begin{eqnarray}
  & & \Mae d_{i}(\dalpha_{\sigma(1)})^{\lambda_{1}}
  (\dalpha_{\sigma(2)})^{\lambda_{2}}\cdots
  (\dalpha_{\sigma(l)})^{\lambda_{l}}
  \cdot 1\nonumber\\
  & & = \bigl[d_{i},(\dalpha_{\sigma(1)})^{\lambda_{1}}
  (\dalpha_{\sigma(2)})^{\lambda_{2}}\cdots
  (\dalpha_{\sigma(l)})^{\lambda_{l}}
  \bigr]\Sym\cdot 1.
  \label{eqn:TP_1}
\end{eqnarray}
>From eq. \myref{eqn:Commutator_harmonic_Dunkl_2}, we 
can easily verify
\begin{eqnarray}
  & & \Mae \bigl[d_{i},\dalpha_{j}\bigr]\nonumber\\
  & & = \dalpha_{i}\Bigl\{\delta_{ij}
  \bigl(1+a\sum^{N}_{\stackrel{\scriptstyle k=1}{k\neq i}}
  K_{ik}\bigr)
  -a(1-\delta_{ij})K_{ij}\Bigr\}.
  \label{eqn:TP_2}
\end{eqnarray}
Using the above formula, we get the following expressions:
\begin{subequations}
  \begin{eqnarray}
    & & \;\Mae \bigl[d_{i},(\dalpha_{\sigma(1)})^{\lambda_{1}}
    (\dalpha_{\sigma(2)})^{\lambda_{2}}\cdots
    (\dalpha_{\sigma(l)})^{\lambda_{l}}
    \bigr]\Sym\nonumber\\
    & & = \Bigl\{\bigl(\lambda_{h} + (N-l)a\bigr)
    (\dalpha_{\sigma(1)})^{\lambda_{1}}
    (\dalpha_{\sigma(2)})^{\lambda_{2}}\cdots
    (\dalpha_{\sigma(l)})^{\lambda_{l}}\nonumber\\
    & & \Cusr + a\sum_{k=1}^{\lambda_{h}-1}
    \sum_{j=l+1}^{N}\nksd{1}{h}{l}\nonumber\\
    & & \Usr (\dalpha_{\sigma(h)})^{k}
    (\dalpha_{\sigma(j)})^{\lambda_{h}-k}\nonumber\\    
    & & \Cusr + a\sum_{j=h+1}^{l}
    \nknkd{1}{h}{j}{l}\nonumber\\
    & & \Usr \sum_{k=1}^{\lambda_{h}-\lambda_{j}}
    (\dalpha_{\sigma(h)})^{\lambda_{h}-k+1}
    (\dalpha_{\sigma(j)})^{\lambda_{j}+k-1}
    \nonumber\\
    & & \Cusr - a\sum_{j=1}^{h-1}
    \nknkd{1}{j}{h}{l}\nonumber\\
    & & \Usr \sum_{k=1}^{\lambda_{j}-\lambda_{h}}
    (\dalpha_{\sigma(h)})^{\lambda_{j}-k+1}
    (\dalpha_{\sigma(j)})^{\lambda_{h}+k-1}\Bigr\}\Sym,
    \nonumber\\
    & & \Usr\Usr\Usr i=\sigma(h),\;1\leq h \leq l,
    \label{eqn:TP_3a}\\
    & & \;\Mae \bigl[d_{i},(\dalpha_{\sigma(1)})^{\lambda_{1}}
    (\dalpha_{\sigma(2)})^{\lambda_{2}}\cdots
    (\dalpha_{\sigma(l)})^{\lambda_{l}}
    \bigr]\Sym\nonumber\\
    & &  = \Bigl\{-a \sum_{k=1}^{l}
    \nksd{1}{k}{l}(\dalpha_{\sigma(h)})^{\lambda_{k}}
    \nonumber\\
    & & \Cusr -a\sum_{k=1}^{l}
    \sum_{m=1}^{\lambda_{k}-1}\nksd{1}{k}{l}\nonumber\\
    & & \Usr 
    (\dalpha_{\sigma(h)})^{\lambda_{k}-m}(\dalpha_{\sigma(k)})^{m}
    \Bigr\}\Sym,
    \nonumber\\
    & & \Usr\Usr\Usr i=\sigma(h),\; l+1\leq h \leq N.
    \label{eqn:TP_3b}
  \end{eqnarray}
  \label{eqn:TP_3}
\end{subequations}
Note that the summation $\sum_{k=s}^{e},\;e<s$ means 
zero, which occurs
in the third and fourth term of eq. \myref{eqn:TP_3a} when
$\lambda_{h}=\lambda_{j}$ or $h=1,l$.
>From the above calculation, we get the following result.
\begin{proposition}
  No higher order monomial in the
  dominance ordering is generated by the action of $d_{i}$ 
  operator on
  a monomial $(\dalpha_{1})^{\lambda_{\sigma(1)}}
  (\dalpha_{2})^{\lambda_{\sigma(2)}}\cdots
  (\dalpha_{N})^{\lambda_{\sigma(N)}}\cdot 1$, where
  $\lambda=\{\lambda_{1}\geq\lambda_{2}\geq
  \cdots\geq\lambda_{N}\geq 0\}$
  and $\sigma\in S_{N}$,
  and the coefficients of the monomials
  are integer coefficient polynomials of $a$.
  The weight of the Young tableaux of the monomials are the same
  as that of the original monomial.
  \label{prop:Triangular_integer_coefficient}
\end{proposition}
This property causes the triangularity
of the Hi-Jack polynomials \myref{eqn:Hi-Jack_triangularity}.
>From the definition of the generator of the Hi-Jack polynomial 
and the above
proposition, we notice the weak form of the third requirement 
of Proposition
\ref{prop:Hi-Jack}.
\begin{proposition}
  \begin{eqnarray}
    & & \Mae C_{\lambda}j_{\lambda}(\vecvar{x};\omega,1/a)
    \nonumber\\
    & & =(b_{N}^{+})^{\lambda_{N}}
    (b_{N-1}^{+})^{\lambda_{N-1}-\lambda_{N}}
    \cdots (b_{1}^{+})^{\lambda_{1}-\lambda_{2}}\cdot 1\nonumber\\
    & & = \sum_{\mu\ledo\lambda}
    {\rm v}_{\lambda\mu}(a)m_{\mu}(\dalpha_{1},\cdots,
    \dalpha_{N})\cdot 1,
    \label{eqn:Weak_triangularity}
  \end{eqnarray}
  where ${\rm v}_{\lambda\mu}(a)$ 
  is an unspecified integer coefficient polynomial of $a$.
  \label{prop:Weak_triangularity}
\end{proposition}
Combining Corollary \ref{cor:Monomial_Dunkl_Monomial_x_expansion}
and Proposition \ref{prop:Weak_triangularity}, we can easily 
confirm the
third requirement of Proposition \ref{prop:Hi-Jack}, i.e., eq.
\myref{eqn:Hi-Jack_triangularity}.
To verify the fourth requirement \myref{eqn:Hi-Jack_normalization},
we have to show 
\begin{equation}
  {\rm v}_{\lambda\lambda}(a)=C_{\lambda}.
  \label{eqn:Normalization_check}
\end{equation}
Computation of ${\rm v}_{\lambda\lambda}(a)$ 
needs consideration on the
cancellation among the monomials in eq. \myref{eqn:TP_3}.

After summing over the distinct permutation $\sigma\in S_{N}$,
the terms that come from the third and the fourth terms of
\myref{eqn:TP_3a} cancel out and vanish, for the fourth term with
the permutation 
$\sigma$ replaced by $\sigma^{\prime}=\sigma(jh)$ yields
\begin{eqnarray}
  & & \Mae - a\nknkdp{1}{h}{j}{l}\nonumber\\
  & & \sum_{k=1}^{\lambda_{h}-\lambda_{j}}
  (\dalpha_{\sigma^{\prime}(j)})^{\lambda_{h}-k+1}
  (\dalpha_{\sigma^{\prime}(h)})^{\lambda_{j}+k-1}\nonumber\\
  & &  = -a\nknkd{1}{h}{j}{l}\nonumber\\
  & & \Usr \sum_{k=1}^{\lambda_{h}-\lambda_{j}}
  (\dalpha_{\sigma(h)})^{\lambda_{h}-k+1}
  (\dalpha_{\sigma(j)})^{\lambda_{j}+k-1},
  \label{eqn:TP_5}
\end{eqnarray}
which is the summand in the third term of eq. \myref{eqn:TP_3a}
with negative sign.
Similar cancellation also occurs between the second term in the
bracket of the first term of eq. \myref{eqn:TP_3a} and 
the first term of
eq. \myref{eqn:TP_3b} with $\sigma$ replaced by
$\sigma^{\prime}=\sigma(kh),\;l+1\leq k\leq N$, where $\sigma$ in the
r.h.s. is that of eq. \myref{eqn:TP_3a}:
\begin{eqnarray}
  & & \Mae\Cmae -a\nksp{1}{h}{l}
  (\dalpha_{\sigma^{\prime}(k)})^{\lambda_{h}}
  \nonumber\\
  & & \Usr = -a(\dalpha_{\sigma(1)})^{\lambda_{1}}
  (\dalpha_{\sigma(2)})^{\lambda_{2}}\cdots
  (\dalpha_{\sigma(l)})^{\lambda_{l}}.
  \label{eqn:TP_6}
\end{eqnarray}
Since there are $N-l$ permutations $\sigma^{\prime}$ that 
yield the monomial
$-a(\dalpha_{\sigma(1)})^{\lambda_{1}}
(\dalpha_{\sigma(2)})^{\lambda_{2}}\cdots
(\dalpha_{\sigma(l)})^{\lambda_{l}}$ through the commutator
\myref{eqn:TP_3b}, we can cancel the second term of 
the first bracket in the
first term of eq. \myref{eqn:TP_3a}.
Thus the coefficient of the term
$(\dalpha_{\sigma(1)})^{\lambda_{1}}\cdots
(\dalpha_{\sigma(l)})^{\lambda_{l}}$ coming from 
the commutator
$\bigl[d_{i},m_{\lambda}(\dalpha_{1},\dalpha_{2},
\cdots,\dalpha_{N})\bigr]\Sym$
is $\lambda_{h}$ where $i=\sigma(h)$.

Since the Hi-Jack polynomial is symmetric with respect to 
the exchange of
indices $1,\cdots,N$, we have only to calculate the coefficient of
$(\dalpha_{1})^{\lambda_{1}}\cdots(\dalpha_{l})^{\lambda_{l}}$.
In the following calculation, we shall omit all the monomials
except for $(\dalpha_{1})^{\lambda_{1}}
\cdots(\dalpha_{l})^{\lambda_{l}}$, namely
the monomial with identity permutation.
Any lower order monomial and the same order monomial with different
permutation are omitted in the expression.
However, we implicitly sum up over the distinct permutations to use
the above cancellation. To know the coefficient of the 
monomial of interest
that is yielded from
$\bp_{l}m_{\lambda}(\dalpha_{1},\cdots,\dalpha_{N})\cdot 1$,
we have only to do the following calculation using
eq. \myref{eqn:TP_3}:
\begin{eqnarray}
  & & \Mae d_{1,\{l,\cdots,1\}}(\dalpha_{1})^{\lambda_{1}}
  \cdots(\dalpha_{l})^{\lambda_{l}}\cdot 1 \nonumber\\
  & & = d_{1,\{l,\cdots,2\}}\Bigl\{\bigl[d_{1},
  (\dalpha_{1})^{\lambda_{1}}
  \cdots(\dalpha_{l})^{\lambda_{l}}\bigr]\Sym\nonumber\\
  & & \;\;
  +la(\dalpha_{1})^{\lambda_{1}}\cdots
  (\dalpha_{l})^{\lambda_{l}}\Bigr\}\cdot 1
  \nonumber\\
  & & = d_{1,\{l,\cdots,2\}}\Bigl\{(\lambda_{1}+la)
  (\dalpha_{1})^{\lambda_{1}}\cdots(\dalpha_{l})^{\lambda_{l}}
  \nonumber\\
  & & \;\; 
  +a(N-l)(\dalpha_{1})^{\lambda_{1}}\cdots(\dalpha_{l})^{\lambda_{l}}
  \nonumber\\
  & & \;\; +a\sum_{\stackrel{\scriptstyle i=2}{\lambda_{i}<
  \lambda_{1}}}^{N}
  (\dalpha_{1})^{\lambda_{1}}\cdots
  (\dalpha_{l})^{\lambda_{l}}\Bigr\}\cdot 1.
  \label{eqn:TP_7}
\end{eqnarray}
Summing over the distinct permutation, we can cancel 
out the second and third
term as has been explained in eqs. \myref{eqn:TP_5} 
and \myref{eqn:TP_6}.
Next, we operate $d_{2}$ on the operand:
\begin{eqnarray}
  & & \Mae d_{1,\{l,\cdots,1\}}
  (\dalpha_{1})^{\lambda_{1}}
  \cdots(\dalpha_{l})^{\lambda_{l}}\cdot 1\nonumber\\
  & & = (\lambda_{1}+la)d_{1,\{l,\cdots,3\}}\nonumber\\
  & & \;\;\Bigl\{(\lambda_{2}+(l-1)a)
  (\dalpha_{1})^{\lambda_{1}}\cdots(\dalpha_{l})^{\lambda_{l}}
  \nonumber\\
  & & \;\; +(N-l)(\dalpha_{1})^{\lambda_{1}}\cdots
  (\dalpha_{l})^{\lambda_{l}}
  \nonumber\\
  & & \;\; - a\sum_{\stackrel{\scriptstyle i=1}{\lambda_{i}>
  \lambda_{2}}}^{1}
  (\dalpha_{2})^{\lambda_{i}}(\dalpha_{i})^{\lambda_{2}}
  (\dalpha_{3})^{\lambda_{3}}\cdots(\dalpha_{l})^{\lambda_{l}}
  \nonumber\\
  & & \;\; +a\sum_{\stackrel{\scriptstyle i=3}{\lambda_{i}<
  \lambda_{2}}}^{N}
  (\dalpha_{1})^{\lambda_{1}}\cdots(\dalpha_{l})^{\lambda_{l}}
  \Bigr\}\cdot 1.
  \label{eqn:TP_8}
\end{eqnarray}
Since we have already operated $d_{1}$ on the 
monomial symmetric polynomial
$m_{\lambda}(\dalpha_{1},\cdots,\dalpha_{N})$, 
summation over the distinct
permutation is not anymore invariant under the permutation of
the indices $1,\cdots,N$, but invariant under the 
permutation of indices
$2,\cdots,N$. To cancel out the second and fourth terms, 
we only need
transpositions $(2j)$, $3\leq j\leq N$. The third term, 
which is a monomial
with the same Young tableau and a permutation $(12)$, 
can not be canceled out
because of the break of invariance of summation under 
the permutation involving
the index $1$. However, this monomial can not be 
changed to the monomial
with identity permutation by operating $d_{i}$, $i\geq 3$. 
Thus, we have
\begin{eqnarray}
  & & \Mae d_{1,\{l,\cdots,1\}}
  (\dalpha_{1})^{\lambda_{1}}
  \cdots(\dalpha_{l})^{\lambda_{l}}\cdot 1\nonumber\\
  & & = (\lambda_{1}+la)(\lambda_{2}+(l-1)a)\nonumber\\
  & & \;\;d_{1,\{l,\cdots,3\}}
  (\dalpha_{1})^{\lambda_{1}}\cdots(\dalpha_{l})^{\lambda_{l}}
  \cdot 1.
  \label{eqn:TP_9}
\end{eqnarray}
Repeating analogous calculations, we get
\begin{eqnarray}
  & & \Mae d_{1,\{l,\cdots,1\}}
  (\dalpha_{1})^{\lambda_{1}}
  \cdots(\dalpha_{l})^{\lambda_{l}}\cdot 1\nonumber\\
  & & = (\lambda_{1}+la)(\lambda_{2}+(l-1)a)\cdots(\lambda_{l}+a)
  \nonumber\\
  & & \Cusr (\dalpha_{1})^{\lambda_{1}}\cdots
  (\dalpha_{l})^{\lambda_{l}}
  \cdot 1.
  \label{eqn:TP_10}
\end{eqnarray}
Then the following expansion follows from the above formula,
\begin{eqnarray}
  & & \Mae \bp_{l}m_{\lambda}(\dalpha_{1},\cdots,\dalpha_{N})\cdot 1
  \nonumber\\
  & & = \Bigl\{(\lambda_{1}+la)(\lambda_{2}+(l-1)a)\cdots
  (\lambda_{l}+a)
  \nonumber\\
  & & \Usr
  m_{\lambda+1^{l}}(\dalpha_{1},\cdots,\dalpha_{N})
  \nonumber\\
  & & \Cusr +
  \sum_{\stackrel{\scriptstyle \mu\ledo \lambda+1^{l}}{\mu\neq\lambda}}
  y_{\lambda+1^{l} \mu}(a)m_{\mu}
  (\dalpha_{1},\cdots,\dalpha_{N})\Bigr\}\cdot 1,
  \label{eqn:TP_11}
\end{eqnarray}
where $\lambda+1^{l}=\{\lambda_{1}+1,\cdots,\lambda_{l}+1\}$ and 
$y_{\lambda+1^{l} \mu}(a)$ is an unspecified integer coefficient 
polynomial of $a$.
We remark that $1$ is a monomial symmetric polynomial with 
$\lambda=0$.
Then from the definition of the generator \myref{eqn:Generator} 
and
repeated use of eq. \myref{eqn:TP_11},
we finally verify eq. \myref{eqn:Normalization_check}:
\begin{lemma}
  \[
    {\rm v}_{\lambda\lambda}(a)=C_{\lambda}.
  \]
  \label{lemma:Normalization_check}
\end{lemma}
Combining Corollary \ref{cor:Monomial_Dunkl_Monomial_x_expansion},
Proposition \ref{prop:Weak_triangularity} and Lemma
\ref{lemma:Normalization_check}, we can confirm the third and fourth
requirements of Proposition \ref{prop:Hi-Jack}, eqs.
\myref{eqn:Hi-Jack_triangularity} and 
\myref{eqn:Hi-Jack_normalization},
and Proposition \ref{prop:Macdonald-Stanley_Hi-Jack}.

In the limit $\omega\rightarrow\infty$, 
eq. \myref{eqn:Weak_triangularity}
reduces to
\begin{eqnarray}
  & & \!\!\! C_{\lambda}j_{\lambda}(\vecvar{x};
  \omega\rightarrow\infty,1/a)
  \nonumber\\
  & & =(b_{N}^{+})^{\lambda_{N}}
  (b_{N-1}^{+})^{\lambda_{N-1}-\lambda_{N}}
  \cdots (b_{1}^{+})^{\lambda_{1}-\lambda_{2}}\cdot 1
  \Bigr|_{\omega\rightarrow\infty}\nonumber\\
  & & = \sum_{\mu\ledo\lambda}
  {\rm v}_{\lambda\mu}(a)m_{\mu}(x_{1},\cdots,x_{N}).
  \label{eqn:HJ2J_1}
\end{eqnarray}
As has been remarked in \S\ref{sec:preparations}, 
the Dunkl operators for the
Calogero model reduce to corresponding Dunkl operators
\myref{eqn:Correspondence_two_Dunkl}
for the Sutherland model in the limit $\omega\rightarrow\infty$.
Then it is obvious that the Hi-Jack polynomials in the limit
$\omega\rightarrow\infty$ are the Jack polynomials. Thus we have
\begin{equation}
  C_{\lambda}J_{\lambda}(\vecvar{x};1/a)
  = \sum_{\mu\ledo\lambda}
  {\rm v}_{\lambda\mu}(a)m_{\mu}(x_{1},\cdots,x_{N}),
  \label{eqn:HJ2J_2}
\end{equation}
which means the coefficients $v_{\lambda\mu}(a)$ in eq.
\myref{eqn:Jack_triangularity} and ${\rm v}_{\lambda\mu}(a)$ in
\myref{eqn:Weak_triangularity} are essentially same:
\begin{equation}
  {\rm v}_{\lambda\mu}(a)=C_{\lambda}v_{\lambda\mu}(a).
  \label{eqn:HJ2J_3}
\end{equation}
This proves Proposition \ref{prop:Jack_to_Hi-Jack}.

\section{Summary}\label{sec:summary}
We have studied the Hi-Jack symmetric polynomials, which we proposed
in a previous work~\cite{Ujino_6} through their Rodrigues formula
that is an extension of the Rodrigues formula for the Jack symmetric
polynomials discovered by Lapointe and Vinet.~\cite{Lapointe_1}
A proof of the formula based on the algebraic relations among the
Dunkl operators is given. In the consideration of their
normalizations, we have clarified that
expansions of the Hi-Jack symmetric polynomials in terms of
the monomial symmetric polynomials have triangular forms,
as is similar to the Jack symmetric polynomials. We have also 
confirmed that
the Hi-Jack symmetric polynomials exhibits the property 
corresponding to
the weak form of the
Macdonald-Stanley conjecture for the Jack symmetric
polynomials.~\cite{Lapointe_2}
The Hi-Jack symmetric polynomials and the eigenstates
for the Hamiltonian that was algebraically constructed 
through QISM are
related by the transformation between the Jack 
symmetric polynomials and
the power sum symmetric polynomials.

The orthogonal basis provides a very useful tool for the study of
physical quantities in quantum theory.
The orthogonality of the Hi-Jack symmetric polynomials 
is expected to be
important in the exact calculation of the thermodynamic 
quantities
such as the Green functions and the correlation functions, 
as has been
done for the Sutherland model using the properties of the Jack
polynomials.~\cite{Forrester_1,Forrester_2,Lesage_1,Ha_1}
However, construction of the simultaneous eigenstates for 
the first two
conserved operators is not sufficient to conclude that the 
Hi-Jack symmetric
polynomials form the orthogonal basis of the quantum 
Calogero model,
because of the remaining degeneracy in the two
eigenvalues.~\cite{Ujino_5,Ujino_6}
Orthogonality of the Jack polynomials is proved by 
showing that all the 
commuting conserved operators of the Sutherland model
$\{{\cal I}_{k}|k=1,2,\cdots,N\}$ are simultaneously 
diagonalized
by the Jack polynomials.~\cite{Macdonald_1,Macdonald_2}
Considering the correspondence between the Calogero model 
and the Sutherland
model, we can expect that all the conserved operators of 
the Calogero model
$\{I_{k}|k=1,2,\cdots,N\}$ are also diagonalized by the 
Hi-Jack polynomials.
However, a proof of the orthogonality still remains open.

The Dunkl operators for the Calogero and Sutherland models 
are shown to be
related to the Yangian symmetry and the Star-Triangle
relation.~\cite{Bernard_1,Hikami_2} They are also related to 
the generators
of $W$-algebra.~\cite{Ujino_2,Ujino_3,Ujino_4,Bernard_2,Hikami_1}
Moreover, the Jack polynomials is identified with the 
singular vectors of
the $W_{N}$-algebra.~\cite{Awata_1} Relationship between 
the irreducible
representation of the Yangian and the eigenstates of the spin-$1/2$
generalization of the Sutherland model is also claimed.~\cite{Uglov_1}
Thus relationships with the representation theory of Yangian
and $W$-algebra are also interesting. 

Some other extensions are also hopeful. For instance, a $q$-deformation 
of the Hi-Jack polynomials associated with the relativistic Calogero
model~\cite{Ruijsenaars_1,Diejen_1} (``Hidden-Macdonald'' polynomials),
``Hi-Jack'' polynomials associated with root lattices other than 
$A_{N-1}$~\cite{Olshanetsky_1} and the spin generalization of them
sound attractive.
We expect some progresses in these directions in the near future.

\section*{Acknowledgements}
One of the authors (HU) appreciates Research Fellowships of the 
Japan Society
for the Promotion of Science for Young Scientists.

\end{document}